\documentclass[prb,aps,twocolumn,superscriptaddress,showpacs]{revtex4-1}
\usepackage{amsmath,amsfonts,amssymb}
\usepackage{mathtools} 
\usepackage{braket} 
\usepackage{framed} 
\usepackage[pdftex]{graphicx}
\usepackage{bbm}    
\usepackage{color}
\usepackage{multirow} 

\begin{document}

\title{Conductivity of a generic helical liquid}

\author{Nikolaos Kainaris}
\affiliation{Institut f\"ur Nanotechnologie, Karlsruhe Institute of Technology, 76021 Karlsruhe, Germany}
\affiliation{Institut f\"ur Theorie der Kondensierten Materie, Karlsruher Institut f\"ur
  Technologie, 76128 Karlsruhe, Germany}
  
\author{Igor V. Gornyi} 
\affiliation{Institut f\"ur Nanotechnologie, Karlsruhe Institute of
  Technology, 76021 Karlsruhe, Germany}
\affiliation{A.F. Ioffe Physico-Technical Institute, 194021
  St. Petersburg, Russia}

\author{Sam T. Carr} 
\affiliation{School of Physical Sciences, University of Kent, Canterbury CT2  7NH, United Kingdom}

\author{Alexander D. Mirlin}
\affiliation{Institut f\"ur Nanotechnologie, Karlsruhe Institute of Technology,
 76021 Karlsruhe, Germany}
\affiliation{Institut f\"ur Theorie der Kondensierten Materie, Karlsruher Institut f\"ur
  Technologie, 76128 Karlsruhe, Germany}
\affiliation{Petersburg Nuclear Physics Institute,
 188350 St. Petersburg, Russia}

\date{\today}

\begin{abstract}
A quantum spin Hall insulator is a two-dimensional state of matter consisting of an insulating bulk and one-dimensional helical edge states. While these edge states are topologically protected against elastic backscattering in the presence of disorder, interaction-induced inelastic terms may yield a finite conductivity.  By using a kinetic equation approach, we find the backscattering rate $\tau^{-1}$ and the semiclassical conductivity in the regimes of high ($\omega \gg \tau^{-1}$) and low ($\omega \ll \tau^{-1}$) frequency. By comparing the two limits, we find that the parametric dependence of conductivity is described by the Drude formula for the case of a disordered edge. On the other hand, in the clean case where the resistance originates from umklapp interactions, the conductivity takes a non-Drude form with a parametric suppression of scattering in the {\it dc} limit as compared to the {\it ac} case. This behavior is due to the peculiarity of umklapp scattering processes involving necessarily the state at the ``Dirac point". In order to take into account Luttinger liquid effects, we complement the kinetic equation analysis by treating interactions exactly in bosonization and calculating conductivity using the Kubo formula. In this way, we obtain the frequency and temperature dependence of conductivity over a wide range of parameters. We find the temperature and frequency dependence of the transport scattering time in a disordered system as $\tau \sim [\max{(\omega,T)}]^{-2K-2}$, for $K>2/3$ and $\tau \sim [\max{(\omega,T)}]^{-8K+2}$ for $K <2/3$.
\end{abstract}

\pacs{}

\maketitle

\section{Introduction}
\label{sec:introduction}

One of the most fascinating advances of recent years has been the discovery of the plethora of quantum states characterized by a non-trivial topological structure of the single-particle wavefunctions.\cite{Hasan_Kane_2010,Moore_2010} While the role of topology in the quantum Hall effect was recognized in the early eighties,\cite{TKNN_1982} it was not realized until much later that related topological states exist with other symmetries.\cite{Haldane_2004}  In particular, it was shown about ten years ago \cite{Kane_Mele_2005a,Kane_Mele_2005b,Bernevig_Zhang_2006} that a particular kind of topological order known as $Z_2$ topological order is present in the quantum spin Hall effect -- like the regular quantum Hall effect this also occurs in two-dimensional systems but in the absence of a magnetic field (in other words, in systems with time-reversal symmetry).  It was around this time the physics of topological insulators really took off when it was predicted \cite{Bernevig_Hughes_Zhang_2006} and subsequently measured experimentally \cite{koenig7,Roth_et_al_2009} that the quantum spin Hall effect is realized in HgTe/CdTe quantum wells.  This state has since also been predicted \cite{Liu_et_al_2008} and observed\cite{Knez_Du_Sullivan_2010,Knez_Du_Sullivan_2011,Knez_2014} in InAs/GaSb quantum wells.

The most striking feature of these topological insulator states is that while the bulk exhibits a spectral gap, the edges (or surfaces in three-dimensional cases) support metallic (gapless) states with curious properties.  In the case of the conventional quantum Hall effects, these edge modes are chiral, with the chirality determined by the sense of the external magnetic field.  These states show a quantized conductance \cite{Halperin_1982} as the chiral nature implies there are no states to which electrons may be backscattered and hence no mechanism by which electrical resistance may be generated.  The edge states of a quantum spin Hall system however are quite different.  They form a \textit{helical liquid}, meaning that the chirality and spin-polarisation are linked.\cite{Wu_Bernevig_Zhang_2006,Xu_Moore_2006}  In these helical liquids, an electron moving one direction forms a Kramer's doublet with that moving the opposite direction meaning that a time-reversal symmetric impurity (for example a non-magnetic impurity) can not elastically backscatter electrons.  This is a state sometimes dubbed a \textit{symmetry protected topological state};\cite{Gu_Wen_2009} breaking time-reversal symmetry annuls the topological protection (for example, helical liquids may be formed in other contexts\cite{Young_2014,Mazo_et_al_2014} without the role of topology).

While the topological protection forbids \textit{elastic} backscattering from non-magnetic impurities which may naively be thought to lead to quantized conductance $G=G_0=2e^2/h$, no such simple result exists for \textit{inelastic} backscattering when interactions are present.\cite{Xu_Moore_2006,Wu_Bernevig_Zhang_2006}  Many forms of inelastic scattering have been investigated on the side of theory, including multiple scattering off impurities,\cite{Maciejko_et_al_2009,lezmy12} random Rashba spin-orbit coupling,\cite{japaridze10,Crepin_et_al_2012,Geissler_Crepin_Trauzettel_2014} umklapp interactions both with and without impurities,\cite{schmidt12} phonon scattering,\cite{budich12} and scattering from charge puddles.\cite{Vayrynen_Goldstein_Glazman_2013} 

No matter the source of the inelastic scattering, the result of all of these investigations is that the correction to conductance behaves as a power-law in temperature $\delta G \sim T^\alpha$ (or possibly activated behavior in a clean system\cite{schmidt12}), which goes to zero as temperature goes to zero.  The power is dependent on the exact scattering mechanism chosen, as well as Luttinger liquid effects\cite{Gogolin_Book,Giamarchi_Book} which are ever-present in one-dimensional systems.

Experiments on helical edge states have been performed for both short edge channels, where the system length L is much smaller than the mean free path {\it l}, and long edge channels, where $L \gg {\it l }$.
The first experimental study of the temperature dependence of helical edge transport was performed in Ref.~\onlinecite{koenig_PhDthesis_2007,koenig7}. The authors measure the conductance of short edges ($\sim 1 \, \mu m$) of HgTe/CdTe quantum wells and find that it is a quantized close to $G_0$ and depends only weakly on temperature.
Longer edges of the order of $\sim 10-30 \, \mu m$ have been studied in Ref.~\onlinecite{gusev13}. Their results show conductance well below the quantized value and rather temperature independent.
Transport properties have also been measured in InAs/GaSb quantum wells.\cite{Knez_2014} In these systems the measured conductance is close to the quantized value and seems to be insensitive to temperature and even magnetic field variations over a large range of parameters. This behavior is observed for both short and long edge channels.

As possible ways to explain the lack of temperature dependence, a number of potential perturbations that weakly break time-reversal symmetry have been investigated, such as Kondo impurities,\cite{Maciejko_et_al_2009,Tanaka_Furusaki_Matveev_2011} dynamical nuclear polarization,\cite{DelMaestro_Hyard_Rosenow_2013} exciton condensation\cite{Pikulin_Hyart_2014} and explicit addition of a magnetic field.\cite{lezmy12} No consensus has yet been reached to explain the discrepancy between theory and experiment and therefore more work must be done on both sides.

In this work, we concentrate on the time-reversal symmetric case, where we study the transport properties of a model that was first introduced in Ref.~\onlinecite{schmidt12} and includes interactions, impurity scattering, and a Rashba spin-orbit term. In their work the authors study corrections to the {\it dc} conductance of short edges, while we concentrate on the conductivity of long edge channels.  

While the frequency dependence of the conductivity of a Luttinger liquid (LL) has been studied both in the semiclassical regime \cite{Mirlin_Polyakov_Vinokur_2007, Rosenow_Glatz_Natterman_2007} and including weak localisation effects,\cite{Giamarchi_Schulz_1988, Gornyi_Polyakov_Mirlin_2007} the conductivity of a helical Luttinger liquid (HLL) remains largely unexplored.

We study the conductivity of a HLL first by means of a kinetic equation approach, from which we obtain the high and low frequency limits of conductivity.  This fermionic approach is valid for a weakly interacting system when the Luttinger liquid constant $K\approx 1$; to investigate the more general case we supplement this approach with bosonization, being careful to highlight the links between this and the prior conceptually transparent fermionic calculations.  This allows us to build up a complete picture of the conduction properties of this model over the whole of parameter space.






The structure of the paper is as follows.  In section \ref{sec:model}, we introduce the microscopic model of the helical edge state, while in section III we discuss the kinetic equation formalism and the scattering terms which appear in the collision integral.  In section IV, we give the solution of the kinetic equation for both the {\it ac} and {\it dc} limits, providing a critical comparison of these results.  In section V we derive the bosonized version of the Hamiltonian (treating the disorder via the replica trick), the conductivity of this model is then derived in section VI, and we conclude in section VII.  Technical details are relegated to the appendices.  

Throughout the paper we use the conventions $\hbar=k_B=v_F=1$ while performing the calculation and restore $\hbar$ and $v_F$ in key results.

\section{Model for helical fermions}
\label{sec:model}

We consider an infinite one-dimensional system of helical fermions. The electrons feel a density-density interaction and are subject to a nonmagnetic random disorder potential.
The Hamiltonian is thus a sum of three parts: 
\begin{equation}
   H = H_{0} +H_{\text{int}} +H_{\text{imp}}. 
\end{equation} 
Additionally, the strong spin orbit coupling in the bulk which leads to the emergence of the helical edge states also breaks the $SU(2)$ spin rotation symmetry in the edge. The resulting helical liquid with broken $S_z$ symmetry is termed ``generic helical liquid". 
The model we use to describe this generic HLL was first introduced in Ref.~\onlinecite{schmidt12}. To fix our notation and to review the main ideas behind the model we will briefly review its derivation. 
The 1D helical system is translation invariant and momenta k are thus good quantum numbers for the eigenstates. Furthermore, to lowest order in spin orbit coupling, the spin degree of freedom of the excitations is frozen out because each chirality has a well-defined spin direction. The effective low energy theory for the edge excitations is thus that of free spinless fermions,
   \begin{align}
      H_0 = \frac{1}{L} \sum_{k} \sum_{\eta=R,L} \, \eta \, k \,  \psi_{k,\eta}^{\dagger}  \psi_{k,\eta}^{}. \label{H0}
   \end{align}
Here $\psi_{k,\eta}$ are fermionic operators and $\eta= R,L= +,-$ denotes chirality. If we assume that time reversal symmetry holds, Kramer's theorem ensures that for any k there exist two orthogonal eigenstates, created by fermionic operators $\psi_{\eta,k}^{\dagger}$ and $\psi_{\bar{\eta},-k}^{\dagger}$ which are related by time reversal $\mathcal{T} \psi_{k,\eta} = \eta \psi_{-k,\bar{\eta}} $.
   
The interaction and disorder contributions for spinful fermions read as
   \begin{align}
      H_{\text{int}} =& \frac{1}{L} \sum_{kqp} \sum_{\sigma \sigma'} V_q \psi_{k,\sigma}^{\dagger} 
                        \psi_{k-q,\sigma}^{} \psi_{p,\sigma'}^{\dagger} \psi_{p+q,\sigma'}^{}
                        \label{Hint},\\
      H_{\text{imp}} =& \frac{1}{L} \sum_{k,q} \sum_{\sigma}  U_{q} \psi^{\dagger}_{ k,\sigma }
                        \psi_{k-q,\sigma }^{}. \label{Himp}
   \end{align}
Here, $\sigma = \uparrow,\downarrow$ denotes the spin in z-direction.   

However, as mentioned before in a generic helical liquid, spin rotation invariance around the z-direction will be broken by spin-orbit terms either due to structural inversion asymmetry or bulk inversion asymmetry in the bulk of the system. We therefore formulate the problem in the chiral basis $(R,L)$ in which the free Hamiltonian is diagonal. In order to perform this rotation we follow Ref.~\onlinecite{schmidt12} and derive the rotation matrix from symmetry arguments.
The operators $\psi_{k,\sigma}$ of an electron with momentum k and spin projection $\sigma$ along the z-axis are related to the chiral operators $\psi_{k,\eta}$ by a momentum dependent SU(2) matrix $B_k$,
   \begin{equation}
       \begin{pmatrix}\psi_{k,\uparrow} \\ \psi_{k,\downarrow} \end{pmatrix} 
       = B_k \begin{pmatrix}\psi_{k,R} \\ \psi_{k,L} \end{pmatrix}. \label{operator rotation}
   \end{equation}
To preserve fermionic commutation relations the matrix has to be unitary $B_k^{\dagger} B_k = \mathbbm{1}$. Moreover, time reversal invariance entails the symmetry $B_k = B_{-k}$.
Because of these constraints the leading terms in $B_k$ for small $k \ll k_0$ can be written as
   \begin{equation}
      B_k= \begin{pmatrix} 1-\frac{k^4}{2 k_0^4} & -\frac{k^2}{k_0^2} \\
           \frac{k^2}{k_0^2} & 1-\frac{k^4}{2 k_0^4} \end{pmatrix}  .      
   \end{equation}    
$k_0$ is an effective parameter that describes the strength of spin-orbit coupling; in the absence of any spin-orbit coupling we have $k_0 \to \infty$. Physically, it can be interpreted as the inverse length scale on which an electron keeps its spin orientation.

In the following we assume that interaction and impurity potentials are momentum independent, $U_q=U$ and  $V_q=V$. In the case of interactions this is justified if the potential is well screened by external media e.g. external gates. For impurities we make the assumption that the disorder potential is short-ranged in real space.
Performing the rotation Eq.~(\ref{operator rotation}) in Eqs.~(\ref{Hint}) and (\ref{Himp}) we obtain
   \begin{align}
      \begin{split} H_{\text{int}}
      =& \frac{1}{L}\sum_{kqp} \sum_{\eta_1 \eta_2 \eta_3 \eta_4} V_q
         [B_k^{\dagger} B_{k-q}]_{\eta_1,\eta_2}\left[B_p^{\dagger} B_{p+q}\right]_{\eta_3,\eta_4}\\
       & \times  \psi_{k,\eta_1}^{\dagger} \psi_{k-q,\eta_2}^{} \psi_{p,\eta_3}^{\dagger} 
         \psi_{p+q,\eta_4}^{},\label{hintsum}\end{split}\\
      H_{\text{imp}}
      =& \frac{1}{L}\sum_{k,q} \sum_{\eta_1 \eta_2 } U_q [B_k^{\dagger} B_{k-q}]_{\eta_1,\eta_2}  
         \psi_{k,\eta_1}^{\dagger} \psi_{k-q,\eta_2}^{}. \label{himpsum}
   \end{align}   

To lowest order in $k/k_0$ the product of rotations can be written in the form
   \begin{align}
      \left[ B_k^{\dagger} B_p \right]_{\eta,\eta'} = \delta_{\eta,\eta'} + \eta \, \delta_{\bar{\eta},\eta'} 
      \frac{k^2-p^2}{k_0^2} \label{rotationmatrix},
   \end{align}
where we use the notation $\bar{R} = L$ and vice versa. Inserting (\ref{rotationmatrix}) into (\ref{hintsum}) and (\ref{himpsum}) yields the interaction terms

\begin{widetext}
   \begin{align}
   \begin{split} 
       H_1 =&  \frac{V}{k_0^4 L} \sum_{k,p,q,\eta}   \left(k^2-(k-q)^2 \right) \left(p^2-(p+q)^2 \right)
                    \psi_{k,\eta}^{\dagger} \psi_{p,\bar{\eta}}^{\dagger} 
                    \psi_{k-q,\bar{\eta}}^{} \psi_{p+q,\eta}^{} ,    \\
       H_2 =&  \frac{V}{L} \sum_{k,p,q,\eta}  \psi_{k,\eta}^{\dagger} 
               \psi_{p,\bar{\eta}}^{\dagger}  \psi_{p+q,\bar{\eta}}^{} \psi_{k-q,\eta}^{}, \\
       H_3 =&  \frac{V}{k_0^4 L} \sum_{k,p,q,\eta}   \left(k^2-(k-q)^2 \right) \left(p^2-(p+q)^2 \right)    
               \psi_{k,\eta}^{\dagger} \psi_{p,\eta}^{\dagger} \psi_{p+q,\bar{\eta}}^{} \psi_{k-q,\bar{\eta}}^{}  ,\\                               
       H_4 =&  \frac{V}{L} \sum_{k,p,q,\eta} \psi_{k,\eta}^{\dagger} \psi_{p,\eta}^{\dagger}  
               \psi_{p+q,\eta}^{}  \psi_{k-q,\eta} ,\\       
       H_5 =&  -\frac{V}{k_0^2 L} \sum_{k,p,q,\eta} \eta \,  (k^2-p^2)  
                \psi_{k+q,\eta}^{\dagger} \psi_{p+q,\bar{\eta}}^{\dagger}  
                \psi^{}_{p,\eta} \psi^{}_{k,\eta} +h.c.  , \label{model}   \\
       H_{\text{imp}} =& \frac{U}{L} \sum_{k,p} \sum_{\eta} \left( \psi_{k,\eta}^{\dagger} 
                         \psi_{p,\eta} + \eta \, \frac{k^2-p^2}{k_0^2}  
                         \psi_{k,\eta}^{\dagger} \psi_{p,\bar{\eta}}\right). 
   \end{split}                                          
   \end{align}
\end{widetext}

\begin{figure}
      \begin{center}
      \includegraphics[width=0.5\textwidth]{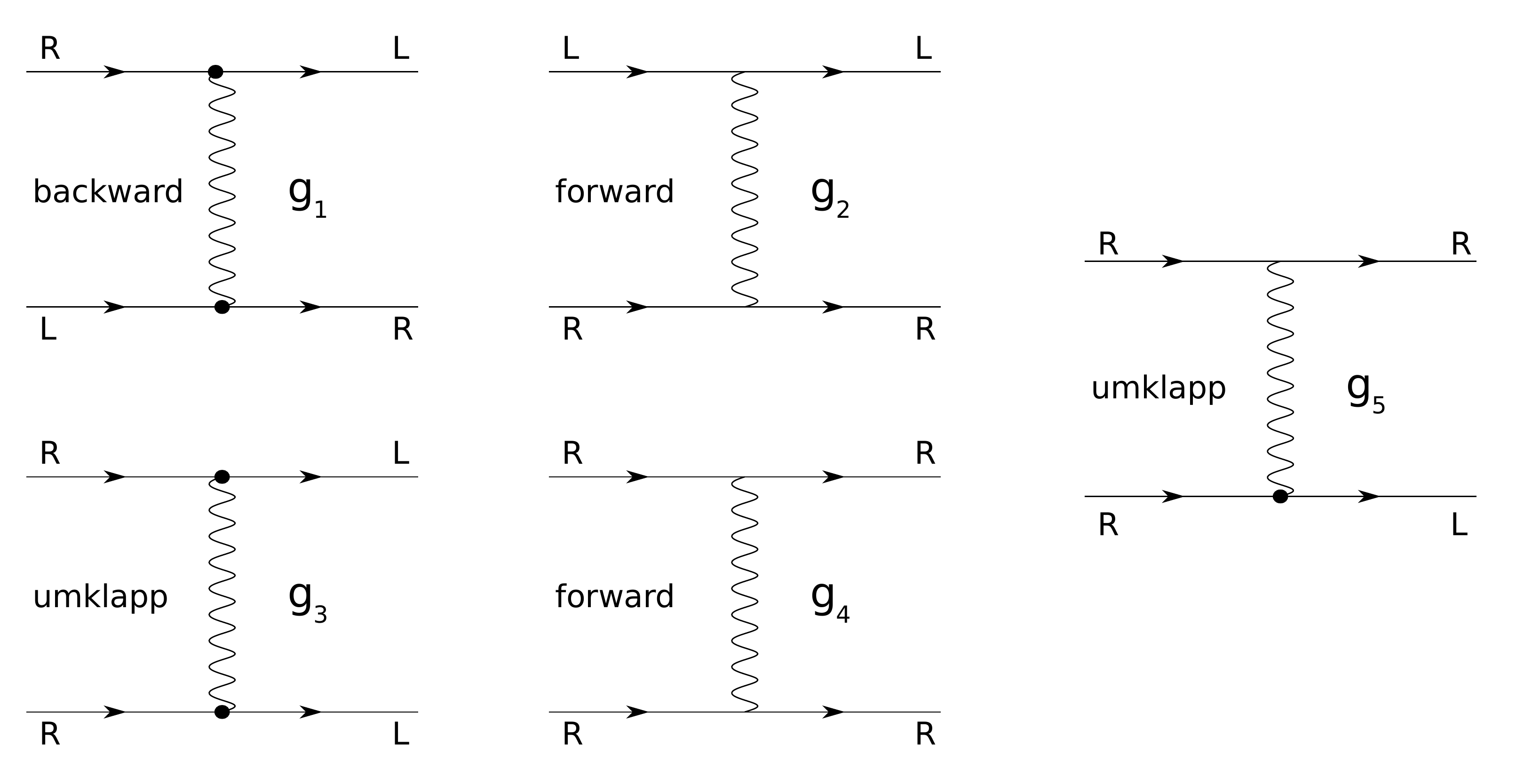}
         \caption{\small G-ology of interaction terms in the HLL.
                         Fat vertices denote chirality changes that have an additional prefactor
                         $\eta_{\text{in}} (k^2_{\text{in}}-k^2_{\text{out}})$.
                         \label{Fig:interactionsHLL}}
      \end{center}   
\end{figure}

The different terms of the interaction Hamiltonian can be grouped analogously to the g-ology of a conventional LL, which motivates our notation. However, in the present model we have an additional umklapp term that backscatters only one incoming particle. For the purpose of this work it will be called $g_5$ term.
A diagrammatic representation of possible interactions processes is depicted in Fig.~\ref{Fig:interactionsHLL}.

It is important to realize, that there is a fundamental difference between the case of conventional one dimensional fermions and the helical fermions discussed here. Usually, one linearizes the spectrum of fermions around the Fermi energy defining left and right movers with linear spectrum. However, both branches of the spectrum are always separated by a large momentum of roughly $2 k_F$. In contrast to that, the spectrum of helical fermions possesses a ``Dirac point", i.e. a point where the right and left moving branches cross. In particular the $g_5$ term only contributes to the low energy physics, if the system is close to the Dirac point which explains why it is never discussed in the context of Luttinger liquid.
However, it will turn out that this process is crucial for the transport properties of a HLL for sufficiently clean samples.

One further thing should be mentioned at this point. The parameters $V$ and $U$ should be considered as effective couplings of the low energy theory after integrating out all degrees of freedom above the UV cutoff, which is given by the bulk gap. Therefore, renormalization effects due to high lying states are already incorporated into the coupling constants and do not affect the physics apart from that.

In the following we investigate the transport properties of this model.
To this end we develop a semiclassical, quantum kinetic equation formalism in the next section.

\section{Quantum kinetic equation formalism}
\label{sec:Quantum kinetic equation formalism}

We assume that the system is subject to some external source of dephasing, such that the dephasing length ${\it l}_{\phi}$ is much shorter than the mean free path {\it l}. In this case we can neglect quantum interference corrections, such as weak localization and describe the system by solving a semiclassical, quantum kinetic equation.

In equilibrium, non-interacting one-dimensional helical fermions have a linear spectrum $\epsilon_{k,\eta} = \eta k$ and obey the Fermi-Dirac distribution $f^{(0)}_{k,\eta} = (1+\exp\{(\epsilon_{k,\eta}-\mu)/T\})^{-1}$.
Away from equilibrium the distribution function $f_{k,\eta}(x,t)$ has to be determined as the solution of a quantum kinetic equation:
   \begin{align}
      \begin{split} & \partial_t f_{k,\eta}(x,t) +v_{k,\eta} \partial_x f_{k,\eta}(x,t) 
                      -eE \partial_k f_{k,\eta}(x,t) \\
                   =& \enspace  I_{k,\eta}[f_{k,\eta}]. \label{KE} \end{split}
   \end{align}
Here, $I_k[f_{\eta}]$ denotes the collision integral which contains all the information about specific scattering processes.

In an infinite, homogeneous wire, we can neglect the spatial dependence of the distribution function.
Furthermore, in linear response to a weak external electric field the electronic distribution function will not differ significantly from the equilibrium Fermi-Dirac distribution and we can expand it as $f_{\eta} \simeq f^{(0)}_{\eta}+f^{(1)}_{\eta} $.
It will prove useful to parametrize the deviation $f^{(1)}$ with another function $\psi$ as
   \begin{align}
      f^{(1)}_{k,\eta}(t) \equiv f^{(0)}_{k,\eta}(1-f^{(0)}_{k,\eta}) \psi_{k,\eta}(t) \label{expansionforf}.
   \end{align}  
Inserting this expansion in our kinetic equation (\ref{KE}) we arrive at the following equation for $\psi$ in the frequency domain
   \begin{align}
        -i\omega \psi_{k,\eta}(\omega)f^{(0)}_{k,\eta}(1-f^{(0)}_{k,\eta}) -eE \partial_k f^{(0)}_{k,\eta}
      = I_{k,\eta}[\psi_{k,\eta}]. \label{KE2}       
   \end{align}
Here, we already made use of the fact that the collision integral is a linear functional and is annihilated by the Fermi distribution i.e. $I_{\eta,k}[f^{(0)}]=0$.\\
Equation~(\ref{KE2}) can formally be rewritten into an integral equation for $\psi$
   \begin{equation}
      \psi_{k,\eta}(\omega) = \frac{I_{k,\eta}[\psi_{k,\eta}]}{(-i\omega) f^{(0)}_{k,\eta}(1-f^{(0)}_{k,\eta})} 
                            - \frac{e E \eta }{(-i \omega ) T} ,\label{psi}
   \end{equation}   
where we used the fact that $\partial_k f^{(0)}_{k,\eta} = - \eta f^{(0)}_{k,\eta} (1-f^{(0)}_{k,\eta})/T $. 
For two particle scattering the collision integral reads as
   \begin{align}
        \begin{split} I_{1}[f_1] =& -  \sum_{2,1',2'} W_{12,1'2'} [f_1 f_2(1-f_{1'})(1-f_{2'})\\
                                & -f_{1'} f_{2'} (1-f_{1})(1-f_{2}) ]. \end{split}
   \end{align} 
Here, we introduced the joint index $1 \equiv (k_1,\eta_1)$.
Since $\psi$ is linear in the electric field and we are interested only in the first order response we can linearize the collision integral in $\psi$: 
   \begin{align}
         \begin{split} I_{1}[\psi_1] 
     =& -\sum_{2,1',2'}  W_{12,1'2'} \Bigl[ f^{(0)}_1 f^{(0)}_2 (1-f^{(0)}_{1'})(1-f^{(0)}_{2'}) \\
      & \times  \left(   \psi_1+\psi_2-\psi_{1'}-\psi_{2'} \right)\Bigr] .\label{twoparticlecollisionintegral}   \end{split}                    
   \end{align}  
The transition probability $W_{12,1'2'}$ is given by Fermi's golden rule
   \begin{equation}
      W_{12,1'2'}= 2\pi \left|\bra{ 1'2'}T \ket{12} \right|^2 \delta(\epsilon_i-\epsilon_f).   \label{transitionprobability}   
   \end{equation}  
The energies in the initial and final states are given by $\epsilon_i = \epsilon_1+\epsilon_2$ and $\epsilon_f = \epsilon_{1'}+\epsilon_{2'}$ and the states $\ket{12}$, $\ket{1'2'}$  are eigenstates of the non-interacting Hamiltonian.     
The T-matrix is given by the expression:
   \begin{align}
   \begin{split}
       T =&    \left(  H_{\text{imp}}+H_{\text{int}} \right) \\
          & + \left(H_{\text{imp}}+H_{\text{int}} \right) G_0
              \left(  H_{\text{imp}}+H_{\text{int}} \right)  + \cdots \, . \label{Tmatrix}
   \end{split}           
   \end{align} 
Here, the Green's function operator is defined as
   \begin{align}
       G_0 = \frac{1}{\eta_i k_i-H_0+i \delta}, \quad \delta \to 0+.
    \end{align}
Some remarks are in order. First, we consider only weak interaction strength $V$ and impurity potential $U$. Therefore, we can restrict our calculation to the lowest orders of the T-matrix.
Second, we assume that impurity scatterers are uncorrelated and therefore the transition probability of a process containing disorder scattering is given by the single impurity probability times the number of impurities.
This is valid as long as the impurity scattering rate is much smaller than the typical electronic energy.\\
Continuing with our formal manipulations let us rename
   \begin{align}
   \begin{split} 
     \Gamma_{12,1'2'} \equiv  & \Bigl[f^{(0)}_1 f^{(0)}_2 (1-f^{(0)}_{1'})(1-f^{(0)}_{2'}) \\
                              & \times \left( \psi_1 + \psi_2 - \psi_{1'} - \psi_{2'} \right) \Bigr]\\
                              & \times  \delta(\eta_1 k_1 + \eta_2 k_2 - \eta_{1'} k_{1'} -\eta_{2'} k_{2'}).  \label{Gamma} 
   \end{split}                                                   
   \end{align}     
Therefore, the final form of the collision integral (\ref{twoparticlecollisionintegral}) reads as
   \begin{align}
          I_{1}[\psi_1]    =& - 2 \pi \sum_{2,1',2'}  \Gamma_{12,1'2'} 
                            \left|\bra{ 1'2'} T \ket{12} \right|^2. \label{collisionintegral2P}
   \end{align} 

\begin{figure}
      \begin{center}
      \includegraphics[width=0.47\textwidth]{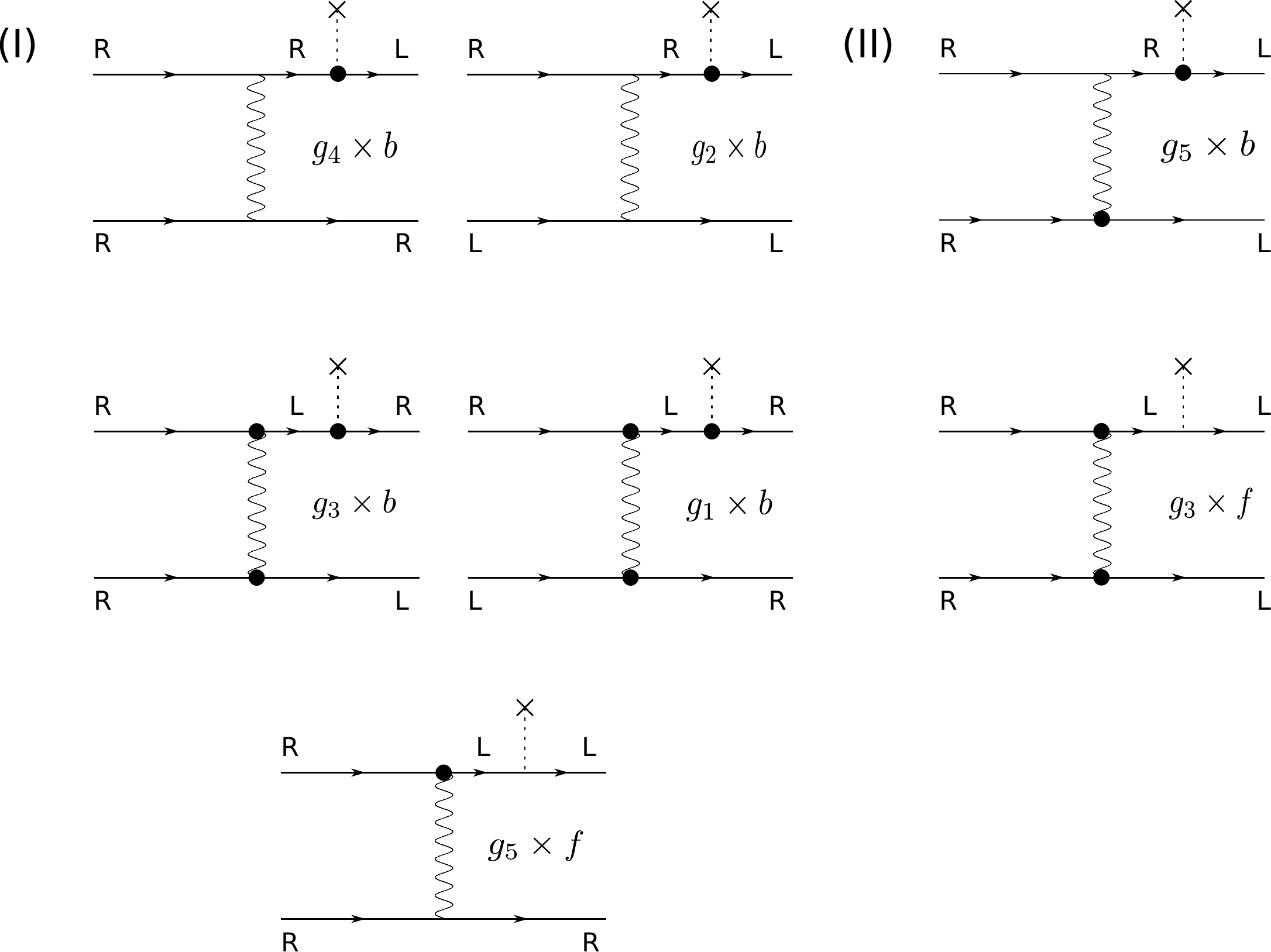}
         \caption{\small Class of interaction processes taken into account as the microscopic realization
                         of inelastic processes backscattering one electron, called "1P processes" (I) and inelastic processes backscattering
                         two electrons, called "2P processes" (II). We study various combinations 
                         of the interaction processes depicted in Fig.~\ref{Fig:interactionsHLL} in combination with backward scattering (b) 
                         or forward scattering (f) off the impurity.
         \label{Fig:perturbationtheory}      }
      \end{center}   
\end{figure}     

After we get the electronic distribution function $f_{k,\eta}$ as the solution of the kinetic equation, we obtain the conductivity as
   \begin{align}
      \sigma =&  -\frac{e}{E L} \sum_{k, \eta} v_{k,\eta} f_{k,\eta} \nonumber\\
                 \stackrel{(\ref{expansionforf})}{=}&
                 -\frac{e}{E L} \sum_{k, \eta} \eta f^{(0)}_{k,\eta} (1-f^{(0)}_{k,\eta}) \psi_{k, \eta}
                 \label{sigmaDC1}\\
                 \stackrel{(\ref{psi})}{=}&
                  \frac{2 e^2}{h} \frac{1}{(-i \omega)}
                 -\frac{e}{E L } \frac{1}{(-i \omega)}\sum_{k, \eta} \eta I_{k, \eta}[\psi]. \label{sigmaac1}                
   \end{align}
This will be used to calculate the conductivity of weakly interacting fermions in the next section.
   
\section{Conductivity of weakly interacting helical fermions}
\label{sec:Conductivity of helical fermions}

\subsection{Dynamic conductivity}
\label{subsec:Dynamic conductivity}

In the case of frequencies much larger than the inverse transport scattering time,
we can solve the integral equation (\ref{psi}) by iteration: 
   \begin{align}
      \psi_{k,\eta}^{(0)} \equiv& -\frac{e E \eta }{(-i \omega ) T},\\
      \psi_{k,\eta}^{(n+1)} =& \frac{I_{\eta,k}[\psi^{(n)}]}{(-i\omega) f^0_{\eta,k}(1-f^0_{\eta,k})}
                              +\psi_{\eta,k}^{(n)}, \quad n \in \mathbb{N}.       
   \end{align}
Here, we stop at the zeroth order which leads to the conductivity, cf. Eq.~(\ref{sigmaac1}), 
   \begin{align}
       \sigma_{\text{{\it ac}}} =& \frac{2 e^2}{h} \frac{1}{(-i \omega)} 
                           - \frac{e}{E L (-i \omega)}\sum_{k, \eta} \eta
                             I_{k, \eta}[\psi^{(0)}] \label{sigmaAC}.
   \end{align} 
The entire information about specific scattering mechanisms is encoded in the collision integral. In the following we will discuss certain microscopic mechanisms and their impact on transport. In particular, we are interested in the real part of conductivity that arises due to these collisions and characterizes current relaxation.   

To calculate the real part of the conductivity we proceed as follows. First we calculate the matrix elements $\bra{ 1'2'} T \ket{12}$ of the T matrix order by order in the expansion in Eq.~(\ref{Tmatrix}). From these expressions we obtain the collision integral according to Eq.~(\ref{collisionintegral2P}), where now the distribution functions $\psi$ are replaced by the zeroth order approximation $\psi^{(0)}$. The obtained collision integrals are then used to calculate the conductivity as explained in Eq.~(\ref{sigmaAC}).

To first order in the T-matrix we consider interactions and disorder separately, $T = H_{\text{int}} +H_{\text{imp}}$. The conductivity of a clean interacting system is discussed in Sec.~\ref{sec:Dynamic conductivity:subsec:Clean system} and some details of the calculation can be found in Appendix~\ref{App:g5fermionic}. Due to the topological protection of the edge states, disorder does not lead to a finite conductivity by itself. Therefore, we have to consider the second order of the T-matrix expansion where combined effects of interactions and disorder appear as $ \bra{1'2'} H_{\text{int}} G_0 H_{\text{imp}} \ket{12} + \bra{1'2'} H_{\text{imp}} G_0 H_{\text{int}} \ket{12}$. The effect of these contributions on transport is discussed in Sec.~\ref{subsec:Dynamic conductivity:Disordered system} and some details of the calculation can be found in Appendix~\ref{App:Discussion $g_5$ with forward scattering}.
\begin{figure}
      \begin{center}
      \includegraphics[width=0.45\textwidth]{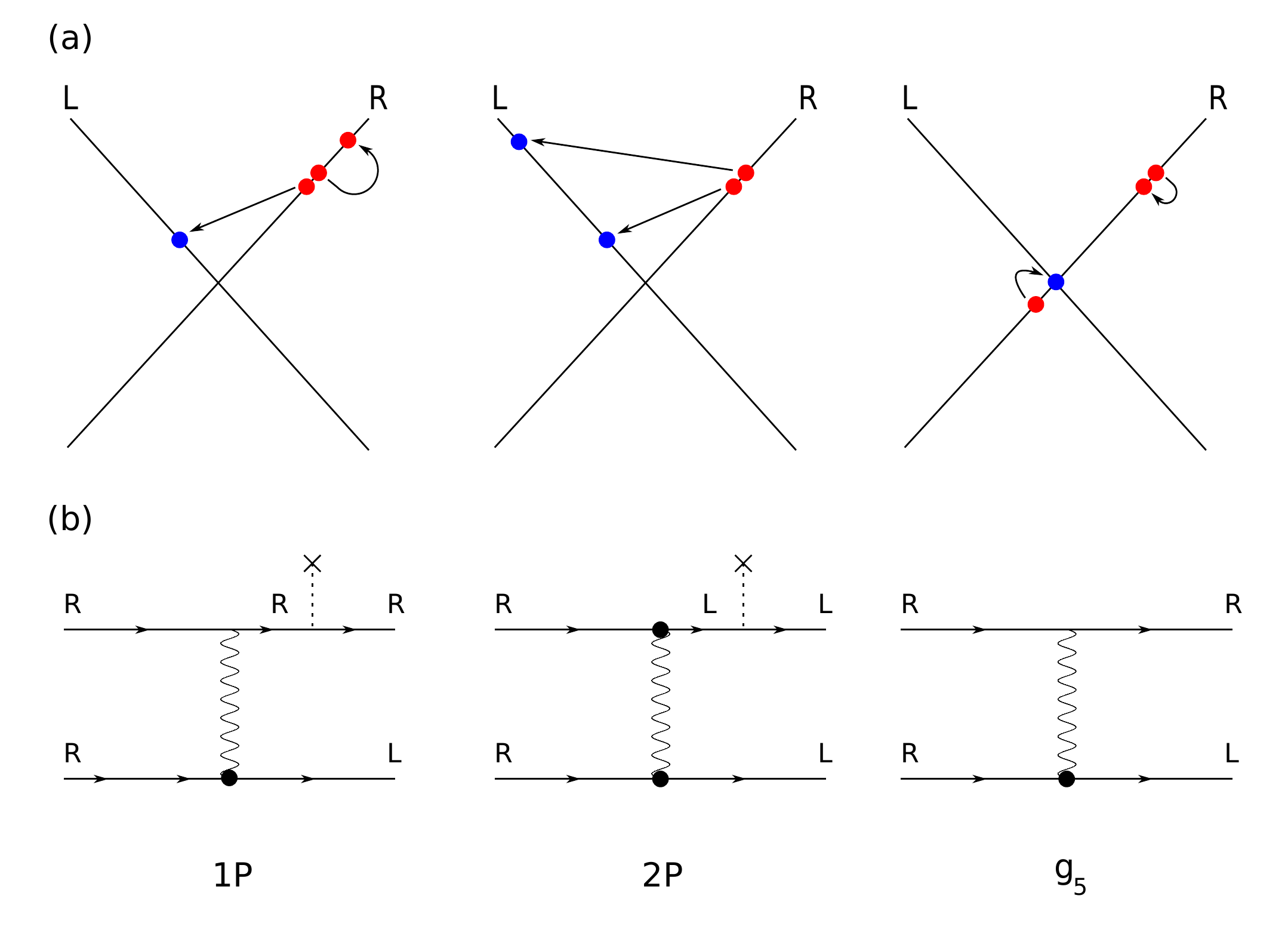}
         \caption{\small Most important scattering processes (a) and their possible microscopic realizations (b).
          While other microscopic combinations of interaction and impurity scattering yield similar outcomes, calculation shows that these combinations
          are the dominant ones (see Tab.~\ref{Tab:results of second order perturbation theory }).
          "1P" describes inelastic processes backscattering one electron and "2P" denotes inelastic processes backscattering two electrons.
          While $g_5$ is a pure interaction effect, the 1P and 2P scattering events also contain impurity scattering. Due to the presence of disorder
          the latter processes do not have to conserve momentum which enlarges the available phase space.
          Therefore, processes containing both interaction and disorder scattering lead to the most important terms in the conductivity. 
         \label{Fig:scatteringprocesses}      }
      \end{center}   
\end{figure}   
   
\subsubsection{Clean system} 
\label{sec:Dynamic conductivity:subsec:Clean system}

In the absence of any impurity scattering we find a finite real part of the conductivity due to
$g_5$ processes, see Appendix~\ref{App:g5fermionic}. The resulting expression reads as
   \begin{align}
       \Re \sigma =& \frac{e^2 v_F }{h} \frac{1}{\omega^2}  \left(\frac{V}{v_F} \right)^2 
                     v_F k_0 \left(\frac{T}{v_F k_0} \right)^5 f(\zeta). \label{referAppA}
   \end{align}
Here, we defined the ratio $\zeta = k_F/T$ and the dimensionless function   
   \begin{align}    
   \begin{split}
      f(\zeta) =& \frac{8}{\pi}  \int \! \mathrm{d} x  \mathrm{d} y \,  (x^2-y^2)^2  n_F(x-\zeta) n_F(y-\zeta)  \\ 
                &\times(1-n_F(x+y-\zeta) )(1-n_F(-\zeta) ),  \label{f(zeta)}
   \end{split} 
   \end{align}
where $n_F(x) = (1+e^{x})^{-1}$ is the Fermi function. 
We can find analytical approximations for $f(\zeta)$ for high and low temperatures compared to the Fermi energy.   
In the regime $k_F \gg T$ we obtain $f(\zeta) \simeq 44/45 \zeta^6 e^{-\zeta}$ and consequently the real part of conductivity, 
   \begin{align}
       \Re \sigma =&   \frac{44}{45 \pi} \frac{ e^2 v_F }{ h} \frac{1}{\omega^2} v_F k_F   
                       \left(\frac{V}{v_F} \right)^2 \left(\frac{k_F}{k_0} \right)^6 \frac{v_F k_F}{T}
                       e^{- v_F k_F/T}.
                       \label{conductivityinteraction1}
   \end{align}    
In this regime the conductivity is thermally activated because energy and momentum conservation constrict one of the particles in the final state to be created at zero momentum deep within the filled Fermi sea (see Fig.~\ref{Fig:scatteringprocesses}).

Conversely, in the high temperature regime $k_F \ll T$ we get $f(\zeta=0) \simeq 306.02$ and the process leads to power law behavior,
   \begin{align}
      \Re \sigma =& 306.02 \,  \frac{e^2 v_F }{h} \frac{1}{\omega^2} v_F k_0  \left(\frac{V}{v_F} \right)^2  
                      \left(\frac{T}{v_F k_0}\right)^5. \label{conductivityinteraction2}
   \end{align} 
Let us adress one important point: how is it possible that interactions that conserve momentum, such as the $g_5$ term, lead to current relaxation? This is surprising since in conventional Fermi liquids translational invariance implies momentum conservation and entails the persistence of currents in the absence of momentum nonconserving interactions such as impurity scattering. However, in the present case we are dealing with an effective low energy theory in which the current of a one-dimensional electron system is determined by the number of left and right movers. In particular, momentum conservation does not imply current conservation. Current relaxation arises from the scattering of right to left movers or vice versa. While these scattering processes conserve quasimomentum in the effective low energy theory they are in fact umklapp processes in the original lattice model.

In summary, we observe that in the present model only scattering processes that change the total number of left and right movers can lead to a finite conductivity.

Consequently, it is clear that $g_1$,$g_2$ and $g_4$ processes will not affect the current since none of them change the number of left and right movers. 
In principle one might expect that the $g_3$ process also influences transport. However, we find that it does not lead to a finite real part of the conductivity. To develop a deeper understanding of the physics behind this, we consider the translation operator $P_T$ and the particle current $J_0$ of the free Hamiltonian Eq.~(\ref{H0}),
\begin{align}
   P_T =&  \frac{1}{L} \sum_{k,\eta} \psi_{k,\eta}^{\dagger} k \psi_{k,\eta}^{}, \\
   J_0 =&  \frac{1}{L} \sum_{k,\eta}  \eta \psi_{k,\eta}^{\dagger}  \psi_{k,\eta}^{}.
\end{align}
For the case of a clean LL, it was first realized in Ref.~\onlinecite{rosch00} that there exists a linear combination $P_0 = P_T + k_F J_0$ that can be identified as the total momentum of the hamiltonian and is therefore conserved, but also commutes with a single umklapp term. The conclusion is, that a single umklapp term in a conventional LL can never lead to a finite conductivity.
In the present case of a HLL, cf. Eq.~(\ref{model}), we find on the one hand $[H_3,P_0] = 0$, but on the other hand
\begin{align}
\begin{split} 
      [P_T,H_5] =&  \frac{2 V}{k_0^2 L} \sum_{k,p,q,\eta} \eta (p-q)(k^2-p^2) \\
                 &  \times \psi_{k+q,\eta}^{\dagger} \psi_{p-q,\bar{\eta}}^{\dagger} \psi^{}_{p,\eta} \psi^{}_{k,\eta}     -h.c. ,   \\
      [J_0,H_5] =&  \frac{2 V}{k_0^2 L} \sum_{k,p,q,\eta} (k^2-p^2) \\
                 &  \times \psi_{k+q,\eta}^{\dagger} \psi_{p-q,\bar{\eta}}^{\dagger} \psi_{p,\eta} \psi_{k,\eta} -h.c. .      
\end{split}                      
\end{align}
Therefore, there exists no such simple conservation law for the $g_5$ term.
Consequently, we expect a finite conductivity due to $g_5$ but not due to $g_3$ umklapp processes, which is exactly the result obtained in the previous calculation. To see how these conservation laws appear in the kinetic equation formalism we show the explicit calculation for $g_3$ and $g_5$ in Appendix~\ref{App:AC conductivity due to Umklapp interaction}.

\subsubsection{Disordered system}
\label{subsec:Dynamic conductivity:Disordered system}

\begin{table}      
    \squeezetable
    \begin{tabular}{||c|c||}
       \toprule 
       \multicolumn{2}{||c||}{$\tau^{-1}$ for processes that backscatter a single electron} \\ \colrule  
       $g_4 \times b$ &  0 \\ \colrule  
       $g_2 \times b$ &  0 \\ \colrule  
       $g_3 \times b$ & $\lambda_1 \frac{2^{11}}{\pi^2} n_{\text{imp}} (U V)^2 
                        \left(\frac{k_F}{k_0} \right)^8 \left( \frac{T}{k_0}\right)^4$ \\ \colrule  
       $g_1 \times b$ & $\lambda_2  \frac{2^7}{\pi^2} n_{\text{imp}} (U V)^2 
                        \left(\frac{k_F}{k_0} \right)^6 \left( \frac{T}{k_0}\right)^6$ \\ \colrule   
       $g_5 \times f$ & $\lambda_1 \frac{2}{\pi^2} n_{\text{imp}}  (U V)^2   \left( \frac{T}{k_0}\right)^4$  \\ \colrule  
       \multicolumn{2}{||c||}{$\tau^{-1}$ for processes that backscatter two electron} \\ \colrule  
       $g_5 \times b$ & 0\\ \colrule 
       $g_3 \times f$ & $\lambda_2 \frac{2^6}{\pi^2}  n_{\text{imp}}  (U V)^2  
                        \left(\frac{k_F}{k_0} \right)^2 \left( \frac{T}{k_0}\right)^6$   \\    \botrule         
      \end{tabular}
      \caption{\small Results of the second order perturbation theory in the T-matrix. $\lambda_1 \simeq 103.9$ and $\lambda_2 \simeq 1757.97$ 
                      are dimensionless integrals defined in Eq.~(\ref{lambda1}) and Eq.~(\ref{lambda2}) in the Appendix.
                      Forward (backward) scattering off impurities is denoted by f (b). The scattering rate is calculated by summing all diagrams
                      that are generated by combining
                      the two said processes cf. Fig.~\ref{Fig:perturbationtheory}.
                      The corresponding {\it ac} conductivity is obtained by Eq.~(\ref{acconductivity}).}  
      \label{Tab:results of second order perturbation theory }                         
\end{table}

We know that pure disorder scattering will not affect transport properties. Indeed, forward scattering does not change the chirality of a particle and elastic backward scattering is prohibited by time reversal symmetry.
However, it turns out that combined scattering mechanisms that include both interaction and disorder can lead to a finite conductivity.

Using the intuition obtained from the first order of perturbation theory we expect that only processes that change the total number of right or left movers can affect current. This is confirmed in the explicit calculation.

We are therefore left with two classes of processes.
First, there are inelastic processes that change the chirality of a single incoming particle which we will refer to as ``1P processes". Second, we have inelastic scattering processes that change the chirality of both incoming particles which we will dub ``2P processes".
The processes as well as possible microscopic realizations are depicted in Fig.~\ref{Fig:scatteringprocesses}.

In order to obtain the real part of conductivity induced by these scattering mechanisms we have to take into account all possible microscopic realizations of the different types. The processes taken into account are depicted in Fig.~\ref{Fig:perturbationtheory} and the corresponding results are summarized in Tab.~\ref{Tab:results of second order perturbation theory }.   

In the case of 1P scattering we find that the leading contribution in the limit of low temperatures $k_F \gg T$ comes from combined processes of $g_5$ and forward scattering off an impurity and yield
   \begin{align}
      \Re \sigma = 42.1 \, \frac{e^2 v_F}{h} \frac{1}{\omega^2}   \left(\frac{U V}{v_F^2} \right)^2
                    v_F n_{\text{imp}} \left(\frac{T}{v_F k_0}\right)^4 .\label{conductivitycombined1}
   \end{align}
The explicit derivation of this result can be found in Appendix~\ref{App:Discussion $g_5$ with forward scattering}.
   
While the combination of $g_3$ and backward scattering off the impurity produces the same temperature dependence, the corresponding scattering time is bigger by a parametrically large factor $(k_0/k_F)^8$, see Tab.~\ref{Tab:results of second order perturbation theory }. 

The 1P processes are similar to pure $g_5$ interaction in the sense that they change only the chirality of one particle. However, unlike the conductivity due to interaction, Eq.~(\ref{conductivityinteraction1}), the result for the combined process Eq.~(\ref{conductivitycombined1}) is not exponentially suppressed in the limit $ k_F \gg T$. The exponential suppression in the clean case is due to the fact that momentum and energy conservation force one of the particles to be at $k=0$ deep within the filled Fermi sea. If we include impurities, momentum conservation is broken and the phase space requirements for the process are relaxed which removes the exponential suppression. 

If we assume that the \textit{ac} conductivity obeys Drude's law,
\begin{align}
   \sigma_{\textit{ac}} = \frac{2 e^2 v_F}{h} \frac{1}{\omega^2} \tau^{-1}, \label{acconductivity}
\end{align}
the whole information about a specific scattering process is contained in the transport scattering time $\tau$. From Eq.~(\ref{conductivitycombined1}) we obtain the scattering time of 1P processes,
\begin{align}
   \tau_{\textit{ac}}^{\text{1P}} = 0.047 \frac{1}{v_F n_{\text{imp}}}  \left(\frac{v_F^2}{ U V} \right)^2 \left(\frac{v_F k_0}{T} \right)^4.
\end{align}
Using the obtained {\it ac} scattering time we can make predictions about other physical quantities relevant for transport. In particular the \textit{dc} conductance of a short edge, i.e. if the system length $L$ is much shorter than the mean free path $l$, can obtained as
\begin{align}
   G \simeq \frac{2 e^2}{h} \left( 1-\frac{L}{l} \right), \label{conductance}
\end{align}
where $l = v_F \tau_{\textit{ac}}$. In the case of 1P scattering this would yield a correction $\delta G$ to quantized conductance which reads as
\begin{align}
   \delta G = 21.1 \frac{e^2}{h} L n_{\text{imp}}   \left(\frac{U V}{v_F^2} \right)^2  \left(\frac{T}{v_F k_0}\right)^4.
\end{align}

This allows us to compare our results to existing work.\cite{schmidt12} There the authors considered the combination of $g_3$ and backward scattering from the impurity. We therefore find a more important microscopic mechanism that leads to a conductance correction larger by a parametrical factor $(k_0/k_F)^8$.

For 2P processes the leading contribution arises from the combination of $g_3$ and forward scattering off the impurity which yields
   \begin{align}
       \Re \sigma =  2.3 \times 10^4 \frac{e^2 v_F}{h} \frac{ v_F n_{\text{imp}}  }{\omega^2} 
                      \left(\frac{U V}{v_F^2} \right)^2  
                      \left(\frac{k_F}{k_0}\right)^2 \left(\frac{T}{v_F k_0}\right)^6 . \label{conductivitycombined2}
   \end{align}
While this process produces subleading corrections in the present case of weakly interacting electrons it will turn out to be the dominant scattering mechanism for $K<2/3$ when we include Luttinger liquid effects in Sec.\ref{sec: Results bosonization}. 

As a general fact we notice that the scattering times originating from microscopic processes containing backward scattering off disorder are always parametrically larger by powers of $k_0/k_F$ compared to those containing forward scattering.

\subsection{{\it dc} conductivity} 
\label{subsec:DC conductivity}

After having discussed the regime of high frequencies we next turn to the opposite limit of {\it dc} conductivity. 
In order to simplify the subsequent calculations we use an effective Hamiltonian derived in Ref.~\onlinecite{lezmy12} for the most relevant scattering mechanisms. These terms would appear in the Hamiltonian under renormalization and describe 1P and 2P scattering processes, respectively. In the previous calculation of the \textit{ac} conductivity we have identified the microscopic origin of these scattering processes and
we fix their coupling constant by demanding that they replicate the results in Eq.~(\ref{conductivitycombined1}) and Eq.~(\ref{conductivitycombined2}).
This yields
   \begin{align}
      H_{\text{1P}}    =&     \frac{ \bar{g}_{1P}}{L^2}  \sum_{k,p,q,q',\eta} k \, \psi_{q',\eta}^{\dagger} 
                       \psi_{q,\bar{\eta}}^{\dagger} \psi^{}_{p,\eta} \psi^{}_{k,\eta}
                       +h.c. .        \label{H1P} \\
      H_{\text{2P}} =& \frac{\bar{g}_{\text{2P}}}{L^2} \sum_{k,p,q,q',\eta} \enspace k q \enspace 
                       \psi_{k,\eta}^{\dagger} \psi_{p,\eta}^{\dagger}
                       \psi^{}_{q,\bar{\eta}}  \psi^{}_{q',\bar{\eta}} ,\label{H2P}      
   \end{align}
with the coupling constants
\begin{equation}
\bar{g}_{1P} = \sqrt{2 n_{\text{imp}}} \frac{U V}{k_0^2} \enspace \text{and} \enspace \bar{g}_{2P} = 8 \sqrt{ n_{\text{imp}}} \frac{U V k_F}{k_0^4}. 
\end{equation}

To study transport behavior in the {\it dc} limit we proceed as follows. Eq.~(\ref{psi}) represents an exact integral equation determining the distribution function $\psi_{k,\eta}$, where the information about the specific scattering process is encoded in the collision integral.
First, we calculate the collision integrals for the process under consideration and insert it into Eq.~(\ref{psi}). Then we perform the limit $\omega \to 0$ to obtain equations determining the distribution function in the \textit{dc} limit. The distribution functions obey a certain symmetry connecting right and left moving particles, see Eq.~(\ref{symmetrypsi}) in Appendix~\ref{App:AC conductivity due to Umklapp interaction}. Therefore, the integral equations for right and left movers decouple and we consider only the integral equations for right movers $\psi_R(x) \equiv \psi(x)$. Subsequently, we solve the integral equations numerically and obtain the {\it dc} conductivity Eq.~(\ref{sigmaDC1}) as
\begin{align} 
   \sigma_{\text{DC}} =& -\frac{2 e}{E h} T \int \! \mathrm{d} x \, n_F(x-\zeta) (1-n_F(x-\zeta)) 
                         \psi_{\zeta}(x) .\label{sigmaDC}              
\end{align}
Here, $x = k/T$, $\zeta = k_F/T$ are dimensionless momenta and $n_F(x)= (1+e^{\beta x})^{-1}$ is the Fermi function.

\subsubsection{Clean system}
\label{sec:DC conductivity:subsec:Clean system}

From our discussion of the {\it ac} conductivity we know that only the $g_5$ term affects transport properties of a clean system.
In the Appendix~\ref{App:Kinetic equation: Calculation of dc conductivity:clean case} we solve the integral equation for the distribution function and obtain the \textit{dc} conductivity. During the calculation we notice the curious fact that the distribution function of the state at the Dirac point explicitely affects the distribution of all other momentum states. This fact will become crucial for the transport properties in the dc limit. 

We find the conductivity in the regime $k_F \ll T$, 
\begin{align}
   \sigma(k_F \ll T) = 0.014 \times  \frac{2 e^2 v_F}{h} \left( \frac{v_F}{V} \right)^2  \frac{1}{v_F k_0}
                   \left( \frac{v_F k_0}{T} \right)^5, \label{referAppA1}
\end{align}
and in the regime $k_F \gg T$,
\begin{align}
   \sigma(k_F \gg T) = 0.81 \times \frac{2 e^2 v_F}{h} \left( \frac{v_F}{V} \right)^2 
                       \left(\frac{k_0}{k_F}\right)^{4} \frac{1}{v_F k_F} e^{\frac{v_F k_F}{T}}.\label{referAppA2}
\end{align} 

If we assume that the results have the form predicted by the Drude formula in the \textit{dc} limit $$\sigma_{\text{{\it dc}}} = \frac{2 e^2 v_F}{h} \tau$$ and extract the corresponding scattering time $\tau$, we can compare the scattering times obtained in the {\it dc} limit with those in the {\it ac} limit in Eq.~(\ref{conductivityinteraction1}) and  Eq.~(\ref{conductivityinteraction2}).
While there is no parametric difference in the regime of high temperatures this is not the case for low temperatures.
To be more specific, in the regime $k_F \gg T$ the scattering time in the {\it ac} limit is parametrically smaller by a factor $T/k_F$ compared to the scattering time in the {\it dc} limit. This is due to the fact that the state at the Dirac point influences all other momentum states and we will further elaborate on this result in the discussion in Sec.~\ref{subsec: Comparison and discussion}.
 
\subsubsection{Disordered system}
\label{sec:DC conductivity:subsec:Disordered system}

We now turn to the disordered case where we consider the effective 1P and 2P processes.
Again referring to Appendix \ref{App:Kinetic equation: Calculation of dc conductivity:disordered case} for further details
we find the {\it dc} conductivity in the presence of impurities as
\begin{align}
   \sigma_{\text{1P}} =& \frac{\kappa_1 }{2} \times \frac{ 2 e^2 v_F}{h} \frac{1}{v_F n_{\text{imp}}}  
                         \left(\frac{v_F^2}{ U V} \right)^2 \left(\frac{v_F k_0}{T} \right)^4, \label{referAppA3}\\
   \sigma_{\text{2P}} =& \frac{\kappa_2}{2^6} \times  \frac{2 e^2 v_F}{h} \frac{1}{v_F n_{\text{imp}}}  
                         \left(\frac{v_F^2}{ U V} \right)^2 \left( \frac{k_0}{k_F}\right)^2
                         \left(\frac{v_F k_0}{T} \right)^6,         \label{referAppA4}             
\end{align}
where $\kappa_1 = -0.46$ and  $\kappa_2 = -0.042$. Notice that the conductivity in the presence of disorder is not sensitive to the ratio of Fermi energy and temperature and we obtain a single scattering time in both limits, $k_F \ll T$ and $k_F \gg T$.

\subsection{Discussion: {\it ac} vs {\it dc} conductivity }
\label{subsec: Comparison and discussion}

\begin{table*}
      \begin{tabular}{||l|c|c|c|c||  } 
       \toprule
       \multirow{2}{*}{} &  \multicolumn{2}{c|}{$\tau$ in the {\it ac} limit} & \multicolumn{2}{c||}{$\tau$ in the {\it dc} limit} \\ \cline{2-5}
               &  $ T \ll k_F$ &   $T \gg k_F$          &     $T \ll k_F$ & $T \gg k_F$  \\ \colrule
         $g_5$   &   $0.16 \left(\frac{ v_F }{V} \right)^2 \left(\frac{ k_0}{k_F} \right)^4 \frac{T}{(v_F k_F)^2} e^{\frac{v_F k_F}{T}}$ & 
         $6.5 \hspace{-.1cm} \times\hspace{-.1cm} 10^{-3} \left(\frac{v_F}{V} \right)^2 \frac{1}{v_F k_0 } \left(\frac{v_F k_0}{T} \right)^5   $  
               & $ 0.81  \left(\frac{v_F}{V} \right)^2 \left(\frac{k_0}{k_F} \right)^4 \frac{1}{v_F k_F}e^{\frac{v_F k_F}{T}} $ &  
               $0.014  \left(\frac{v_F }{V} \right)^2  \frac{1}{v_F k_0 } \left(\frac{v_F k_0}{T} \right)^5$  \\ \colrule
         1P &  \multicolumn{2}{c|}{$ 0.047 \frac{1}{v_F n_{\text{imp}}}  \left(\frac{v_F^2}{ U V} \right)^2 \left(\frac{v_F k_0}{T} \right)^4 $} 
            & \multicolumn{2}{c||}{$ 0.23 \frac{1}{v_F n_{\text{imp}}}  \left(\frac{v_F^2}{ U V} \right)^2 \left(\frac{v_F k_0}{T} \right)^4 $}    \\ \colrule
         2P &  \multicolumn{2}{c|}{$ 8.8 \times  10^{-5} \frac{1}{v_F n_{\text{imp}}}  \left(\frac{v_F^2}{ U V} \right)^2 \left(\frac{k_0}{k_F} \right)^2  \left(\frac{v_F k_0}{T} \right)^6 $} 
            &  \multicolumn{2}{c||}{$ 6.5 \times 10^{-4} \frac{1}{v_F n_{\text{imp}}}  \left(\frac{v_F^2}{ U V} \right)^2 \left(\frac{k_0}{k_F} \right)^2  \left(\frac{v_F k_0}{T} \right)^6 $} \\ \botrule          
      \end{tabular}
      \caption{\small Comparison of the most dominant scattering mechanisms and their respective transport
                      scattering time $\tau$ in different regimes of temperature and frequency.
                      The $g_5$ process is solely due to electron-electron interaction while 1P and 2P processes describe combined effects of disorder 
                      and interaction. 
                      \label{Tab:summaryDC}     }      
\end{table*}

We are now in the position to compare the results for {\it dc} and {\it ac} conductivity summarized in Tab.~\ref{Tab:summaryDC}. 
In the presence of disorder we consider effective 1P and 2P processes which describe combined effects of interaction and forward scattering off the impurity. They lead to transport scattering times which are insensitive to the ratio of Fermi energy and temperature. Furthermore, the parametric dependence of the transport scattering time is identical in the low and high frequency regime. This suggests that, in the presence of disorder, the parametric dependence of the conductivity can be approximated by the Drude formula,
\begin{equation}
   \sigma(\omega) = \frac{2 e^2 v_F}{h} \frac{1}{\tau^{-1} -i \omega}.
\end{equation}
Nevertheless, the numerical prefactors in the {\it ac} and {\it dc} limit differ substantially. Therefore, the overall behavior of conductivity is not exactly Drude-like.
  
We find that the dominant contribution to conductivity is due to 1P processes. They lead to Drude-like behavior of the conductivity, irrespective of doping and with a temperature scaling $\sim T^4$ in the {\it ac} and $\sim T^{-4}$ in the {\it dc} limit, respectively. 

For sufficiently clean systems we have to consider the effect of $g_5$ interactions.
In this case we have to distinguish between the high temperature and low temperature regime. If temperature is much larger than the Fermi energy, the conductivity behaves Drude-like which is expected since the high temperature limit corresponds to the classical regime. For low temperatures however, Pauli blocking of the state at the Dirac point leads a scattering time which is much larger in the {\it dc} limit, by a parametrically big factor $E_F/T$, compared to the {\it ac} case.
Indeed, we saw that all scattering processes have to go through the state at the Dirac point. In the {\it ac} case the state is frequently emptied due to the applied field. In the {\it dc} limit this can only happen due to thermal fluctuations which leads to a suppression in the low temperature case.

Another point to appreciate is, that the {\it dc} conductivity in the absence of impurities is finite.
This is indeed surprising, since the free Hamiltonian of our system is that of a spinless LL, which is integrable and therefore characterized by an infinite number of conservation laws, current being one of them. Therefore, once a current is created by an externally applied bias it should never relax. For a conventional LL this statement remains true even in the presence of $g_3$ interaction which breaks some conservation laws.
However, in the present case we have shown that the $g_5$ term, that is particular to the HLL,\cite{remark1} does lead to a finite conductivity, while the $g_3$ term does not. As discussed in Sec.~\ref{sec:Dynamic conductivity:subsec:Clean system} this is caused by the fact that $g_3$ commutes with the total momentum of the system while $g_5$ does not.

\section{Luttinger Liquid effects: Formalism}
\label{sec:LL}

So far we have discussed transport properties of one-dimensional electrons subject to weak interactions and impurity scattering neglecting LL effects. While intuitively more accessible, the fermionic description often proves insufficient to describe the strongly correlated LL state of one-dimensional fermions.

Therefore, we now complement our fermionic analysis by bosonizing the model which takes $g_2$ and $g_4$ interactions into account exactly. 
In real space the model of free fermions with linear spectrum and interaction-induced forward scattering reads as
\begin{align}
\begin{split} 
       H_0       =& \sum_{\eta} \int \! \mathrm{d} x \, \Psi_{\eta}^{\dagger}(x)
                    (-i \eta \partial_x) \Psi_{\eta}(x),\\       
       H_2 =&  V \sum_{\eta} \int \! \mathrm{d} x \, \Psi_{\eta}^{\dagger} 
               \Psi_{\bar{\eta}}^{\dagger} \Psi_{\bar{\eta}} \Psi_{\eta} , \\
       H_4 =&  V    \sum_{\eta} \int \! \mathrm{d} x \, \Psi_{\eta}^{\dagger}(x)
              \Psi_{\eta}^{\dagger}(x+0) \Psi_{\eta}(x)\Psi_{\eta}(x+0).      
\end{split}                            
\end{align}
Thereby, the field operators $\Psi(x)$ are slowly varying on the scale $k_F^{-1}$.
We use the bosonization convention\cite{Gogolin_Book} 
\begin{equation}
   \Psi_{\eta} = \frac{1}{\sqrt{2 \pi a}} e^{-i\sqrt{4 \pi} \eta \phi_{\eta}},
\end{equation}
where $\phi_{\eta}$ are the chiral bosonic fields and $a$ is the inverse UV cutoff. The bosonic fields are obtained as
\begin{equation}
   \varphi = \phi_R +\phi_L  \enspace \text{and} \enspace \theta = \phi_R -\phi_L . 
\end{equation}
We now switch to an action formalism. The free action is renormalized by $g_2$ and $g_4$ interaction and reads as
\begin{align}
   S_0 =&  \int \! \mathrm{d} x \mathrm{d} \tau \, 
           \left[ \frac{uK}{2} (\partial_x \theta)^2 
          +\frac{u}{2 K} (\partial_x \varphi)^2 +i \partial_x \theta \partial_{\tau} \varphi \right].\label{bosonicfree}
\end{align}       
Here, $K$ denotes the Luttinger liquid parameter which is a measure of the fermionic interaction strength and $u$ is the renormalized Fermi velocity. In terms of the interaction strength $g_2 = g_4 = V$ and the Fermi velocity $v_F$ they are given by the expressions
\begin{align}
   K = \frac{1}{(1+V/\pi v_F)^{1/2}}, \quad u = v_F \, (1+V/\pi v_F)^{1/2}  .
\end{align}
Let us now include interaction and disorder terms and derive an effective low energy action.
As a starting point we consider the Hamiltonian in Eq.~(\ref{model}) and expand momenta around the Fermi points, i.e. we write $k=k'+\eta k_F$ and expand in $|k'| \ll k_F$. We also define 
   \begin{align}
      \psi_{k,\eta} = \psi_{k'+\eta k_F, \eta} \equiv \Psi_{k',\eta}.
   \end{align}
This yields the following interaction-induced umklapp terms:
\begin{align}
\begin{split} 
       H_3 =&  \frac{8 k_F^2 V}{k_0^4 } \sum_{\eta} \int \! \mathrm{d} x \, 
               e^{-i 4 k_F \eta x} \left(\partial_x \Psi_{\eta}^{\dagger}\right) \Psi_{\eta}^{\dagger}
               \left( \partial_x \Psi_{\bar{\eta}} \right) \Psi_{\bar{\eta}}, \\
       H_5 =&  \frac{4 V k_F i }{k_0^2 } \sum_{\eta} \int \! \mathrm{d} x \, e^{i 2 k_F \eta x} 
               \Psi_{\eta}^{\dagger} \Psi_{\bar{\eta}}^{\dagger} \Psi_{\eta} 
               \partial_x \Psi_{\eta} +h.c.  \,.  
        \label{effectivefermionicmodel}
\end{split}                            
\end{align}
Notice that we did not consider $g_1$ terms since they are similar to $g_2$ terms but with additional derivatives making them less relevant in the renormalization group sense. Performing the same expansion for the impurity terms yields
\begin{align}                     
       H_{\text{imp},f} =& \sum_{\eta} \int \! \mathrm{d} x \, U_f(x) \Psi^{\dagger}_\eta(x) \Psi_\eta(x),
                           \\
\begin{split} 
       H_{\text{imp},b} =& \frac{2 k_F }{ k_0^2} \int \! \mathrm{d} x \, U_b(x) 
                           \left[ \partial_x \Psi_R^{\dagger} \Psi_L - \Psi_R^{\dagger} 
                           \partial_x \Psi_L \right] \\
                         &  + h.c.    .  
\end{split}                           
\end{align} 
Here, we defined the forward and backward scattering impurity potentials as
   \begin{align}
      U_f(x) = \frac{1}{L} \sum_q U_q e^{i q x},\\
      U_b(x) = \frac{i}{L} \sum_q U_{q+2 k_F } e^{iqx}.
   \end{align}    
   
We consider weak, gaussian correlated disorder
   \begin{align}
      \begin{split} \overline{U(x)} =& 0 ,\\
      \overline{U(x) U(x')} =& D \delta(x-x') ,\end{split}
   \end{align}
where $D= n_{\text{imp}} U^2 $ denotes the disorder strength. 
One can then show that the forward and backward scattering potentials obey
   \begin{align}
      \overline{U_f(x) U_f(x')} = \overline{U_b(x) U_b^{\ast}(x')} = D \, \delta(x-x') .     
   \end{align}

We proceed by bosonizing the model and switching to an action formalism. This yields
\begin{align}
\begin{split}
   S_3 =& \frac{4 k_F^2 V}{\pi^2 a^4 k_0^4 } \int \! \mathrm{d} x \mathrm{d} \tau \, \cos(2 \sqrt{4 \pi} \varphi -4k_F x),\\
   S_5 =&  \frac{8 V}{\sqrt{\pi} a k_0^2 }\int \! \mathrm{d} \tau \mathrm{d} x \, (3 (\partial_x \varphi)^2
            \partial_x \theta + (\partial_x \theta)^3) \\
         &  \times \cos\left( \sqrt{4 \pi} \varphi(x,\tau) -2 k_F x \right) \\
         & +\frac{4 V}{ \pi a k_0^2 }   \int \! \mathrm{d} \tau \mathrm{d} x \, 
            (\partial_x^2 \varphi \partial_x \theta  + \partial_x \varphi \partial_x^2 \theta) \\
         &  \times  \sin\left( \sqrt{4 \pi} \varphi(x,\tau) -2 k_F x \right), \\
   S_{\text{imp},f} =& -\frac{1}{\sqrt{\pi}} \int \! \mathrm{d} x \mathrm{d} \tau \, U_f(x) \partial_x \varphi ,\\
   S_{\text{imp},b} =& \frac{2 i k_F}{\sqrt{\pi} a k_0^2}  \int \! \mathrm{d} x \mathrm{d} \tau \, U_b(x) 
                       \partial_x \theta e^{i \sqrt{4 \pi} \varphi} +h.c. .\end{split} \label{effectiveaction}
   \end{align}
   
In the following we distinguish the clean and the disordered case.
Recall that the fermionic treatment in Sec.~\ref{sec:Dynamic conductivity:subsec:Clean system} lead us to the conclusion that $g_3$ umklapp scattering does not produce a finite conductivity. Therefore, we only consider $g_5$ umklapp interaction in the clean limit. 

In the disordered case, we will derive an effective action containing 1P and 2P processes by averaging over disorder. The models we use in each situation are discussed in Sec.~\ref{bosonic model clean} and \ref{bosonic model disordered}.
Subsequently, the high frequency conductivity of each model is calculated in Sec.~\ref{subsec: Linear response Kubo formalism} using the linear response Kubo formula.

\subsection{Model in the clean case}
\label{bosonic model clean}

While the straigthforward calculation of the conductivity due to the $g_5$ term in Eq.~(\ref{effectiveaction}) is possible, it is difficult due to various reasons. 
Therefore, we restrict ourselves to the relevant case of low energy physics where $\omega,T \ll k_F$ and bosonize the effective form of the $g_5$ term close to the Fermi energy in Eq.~(\ref{effectivefermionicmodel}). This yields
\begin{align}
\begin{split} 
   S_5 =&  \frac{ 4 V k_F }{ \pi^{\frac{3}{2}} a k_0^2 }  \int \! \mathrm{d} x \,\partial_x^2 \theta
           \sin(\sqrt{4 \pi} \varphi-  2 k_F x) \\
        &  -\frac{ 16 V k_F^2 }{ \pi^{\frac{3}{2}} a k_0^2 }  \int \! \mathrm{d} x \, 
           \partial_x \theta  \cos(\sqrt{4 \pi} \varphi-  2 k_F x).\label{bosonicmodelclean}
\end{split}           
\end{align}

\subsection{Model in the disordered case}
\label{bosonic model disordered}

In the fermionic description we noticed that the combined effect of forward scattering off disorder and interaction leads to the dominant effects. Therefore, we proceed by gauging out forward scattering from impurities in Eq.~(\ref{effectiveaction}) using the gauge transformation
   \begin{align}
      \varphi \to \varphi + \frac{K}{u \sqrt{\pi}}\int_{x_0}^x \! \mathrm{d} y \, U_f(y),
      \quad x_0 \to -\infty. \label{gaugeoutforward}
   \end{align}
In order to perform the disorder average we introduce replicas and then average over forward and backward scattering.
The technical details can be found in the Appendix~\ref{App:Average over forward scattering in the bosonic action}.

From now on we use subscripts $D_b \equiv D$ and $D_f \equiv D$ in order to differentiate between the two physically distinct disorder scattering mechanisms.    
 
After averaging over disorder we obtain an effective action local in space but nonlocal in imaginary time where the momentum cutoff of our theory is now given by $D_f \ll k_F$,
\begin{widetext}
   \begin{align}
    \begin{split} 
      S_{\text{2P}} =& - g_{\text{2P}} \sum_{a,b} \int \! \mathrm{d} x \mathrm{d} \tau \mathrm{d} \tau' \, 
                       \cos\left\lbrace 2 \sqrt{4 \pi} \left[ \varphi_a(x,\tau)-\varphi_b(x,\tau') \right] \right\rbrace, \\
      S_{1P} =& -g_{\text{1P},1} \sum_{a,b} \int \! \mathrm{d} x \mathrm{d} \tau \mathrm{d} \tau' \, 
                 \partial_x^2 \theta_a(x,\tau) \partial_{x}^2 \theta_b(x,\tau')  \cos\left\lbrace  \sqrt{4 \pi} 
                 \left[ \varphi_a(x,\tau)-\varphi_b(x,\tau') \right] \right\rbrace \\
              &+g_{\text{1P},2} \sum_{a,b} \int \! \mathrm{d} x \mathrm{d} \tau \mathrm{d} \tau' \,
                 \partial^2_x \theta_a(x,\tau) \partial_{x} \theta_b(x,\tau') 
                 \sin\left\lbrace  \sqrt{4 \pi}  \left[ \varphi_a(x,\tau)-\varphi_b(x,\tau') \right] \right\rbrace,\\                       
      S_{\text{imp,b}} =& -g_{\text{b}} \sum_{a,b} \int \! \mathrm{d} x 
               \mathrm{d} \tau \mathrm{d} \tau' \, \partial_x \theta_a(x,\tau) \partial_x \theta_b(x,\tau')
                      \cos\left\lbrace  \sqrt{4 \pi}\left[ \varphi_a(x,\tau)-\varphi_b(x,\tau') \right] \right\rbrace . \label{bosonicmodel}
     \end{split}
   \end{align}
\end{widetext}
Here, $a,b \in \left\lbrace1, \cdots R\right\rbrace$ are replica indices and R is the number of replicas.
The first two terms correspond to the 1P and 2P processes discussed in 
Sec.~\ref{subsec:Dynamic conductivity:Disordered system}. They originate from $g_5$ and $g_3$ umklapp processes, respectively, in combination with forward scattering off impurities. The last term describes the disorder averaged backscattering off disorder. 
The coupling constants are given by
   \begin{align}
       g_{\text{2P}}   =& \frac{ 2 V^2 k_F^2 K^2 D_f}{\pi^6 a^8 k_0^8 u^2},\\
       g_{\text{1P},1} =& \frac{2 V^2 K^2 D_f}{\pi^5 a^2 k_0^4 u^2}, \\
       g_{\text{1P},2} =& \frac{8 k_F V^2 K^2 D_f}{\pi^5 a^2 k_0^4 u^2 },\\
       g_{\text{imp,b}}    =& \frac{4 D_b k_F^2}{\pi a^2 k_0^4} 
                          +\frac{32 V^2 K^2 k_F^2 D_f}{\pi^5 a^2 k_0^4 u^2}.
   \end{align}
Recall, that elastic backscattering off disorder does not affect transport properties in a HLL. Thus, the term $S_{\text{imp,b}}$ should not lead to a finite conductivity at zero interaction strength, i.e. at $K=1$, even if the coupling constant is nonvanishing in this limit. We will return to this point at a later stage.

At the end of any calculation in the replica formalism, one has to analytically continue the result to R=0. In particular the expectation value of some functional $\mathcal{O}$ of fields $\theta$ and $\varphi$ is obtained as
\begin{equation}
   \braket{\mathcal{O}} = \lim_{R \to 0} \sum_{a=1}^{R} \braket{\mathcal{O}(\varphi_a,\theta_a)}. 
\end{equation}
Details of the replica limit can be found in Appendix~\ref{App:Calculation of the conductivity of a disordered helical Luttinger liquid}.

\subsection{Linear response Kubo formalism}
\label{subsec: Linear response Kubo formalism}

In the presence of an electromagnetic field we couple the vector potential to the canonical momentum via the minimal substitution $\partial_x \theta \to \partial_x \theta + \frac{e}{\sqrt{\pi}} A$.\cite{Gogolin_Book} 
The current is then obtained by varying the action with respect to the vector potential 
\begin{align}
   j = \delta S/\delta A|_{A=0}
\end{align}
and the diamagnetic susceptibility is obtained as 
\begin{align}
      \chi^{\text{dia}}(x-x',\tau-\tau') =
    - \left. \braket{\frac{\delta S}{\delta A(x,\tau) \delta A(x',\tau')}}\right|_{A=0}.
\end{align}
However, notice that the vector potential does not only couple to the free action but also to the perturbations in Eq.~(\ref{bosonicmodelclean}) and Eq.~(\ref{bosonicmodel}). Therefore, we get additional contributions to the current and the diamagnetic suceptibility.  
We will refer to the contributions obtained from the free action, Eq.~(\ref{bosonicfree}), as normal and the contributions linear in coupling strengths as anomalous.
The normal current is 
\begin{equation}
   j_0(x,\tau) = \frac{e K u}{ \sqrt{\pi}} \partial_x \theta(x,\tau)
\end{equation}
and the diamagnetic susceptibility is given by 
\begin{equation}
   \chi_0^{\text{dia}}(x-x',\tau-\tau') =  \frac{e^2 u K}{\pi} \delta(x-x') \delta(\tau-\tau').
\end{equation}

In Appendix~\ref{App:Formalism for conductivity calculation in the bosonized language} we state the anomalous part of the current and diamagnetic susceptibility. The total current is then $j = j_0 +j_{\text{an}}$ and analogously $\chi^{\text{dia}}(x,\tau) = \chi_0^{\text{dia}}(x,\tau) +\chi_{\text{an}}^{\text{dia}}(x,\tau)$.

From this we obtain the susceptibility and the conductivity in linear response
   \begin{align}
      \begin{split} \chi(x,\tau) =& \chi^{\text{dia}}(x,\tau) + \braket{j(x,\tau) j(0,0)} , \end{split}\\
      \sigma(\omega,T)      =& -\frac{i}{\omega} \chi(k\to 0, i k_n \to \omega + i \delta), \enspace \delta= 0+.
   \end{align}   
This procedure yields the {\it ac} conductivity of a free system:
\begin{align}
   \sigma_0(\omega) =  \frac{2 e^2 }{h}\frac{i u K}{\omega+i \delta} . \label{sigma0}
\end{align}
To obtain a finite real part of the conductivity we perform a perturbative expansion of the current-current correlator to the lowest nontrivial order in the considered scattering mechanism which is discussed in the following section.

\section{Conductivity of a helical Luttinger liquid for arbitrary interaction stregth}
\label{sec: Results bosonization}

In the following we calculate the conductivity of a HLL for arbitrary interaction stregths , when Luttinger liquid renormalization effects are crucially important. In order to treat the effect of scattering processes pertubatively the corresponding scattering rate has to be the lowest energy scale in the problem. In particular we have to require $ \omega \gg \tau^{-1}$ which means the pertubative treatment only allows us to calculate the {\it ac} conductivity.

The {\it ac} conductivity in the clean case is discussed in Sec.~\ref{sec:bosonization:subsec:clean case}. Sec.~\ref{sec:bosonization:subsec:disordered case} is devoted to the conductivity in the presence of disorder and in Sec.~\ref{sec:bosonization:subsec:discussion} we then discuss the implication of these results on transport in the {\it dc} limit and localization effects. 
Some details of the calculation are summarized in the Appendices \ref{App:Formalism for conductivity calculation in the bosonized language} to \ref{App:Calculation of the conductivity of a disordered helical Luttinger liquid}. 

\subsection{{\it ac} transport in the clean case}
\label{sec:bosonization:subsec:clean case}

\begin{figure}
\begin{center}
   \includegraphics[width=0.5\textwidth]{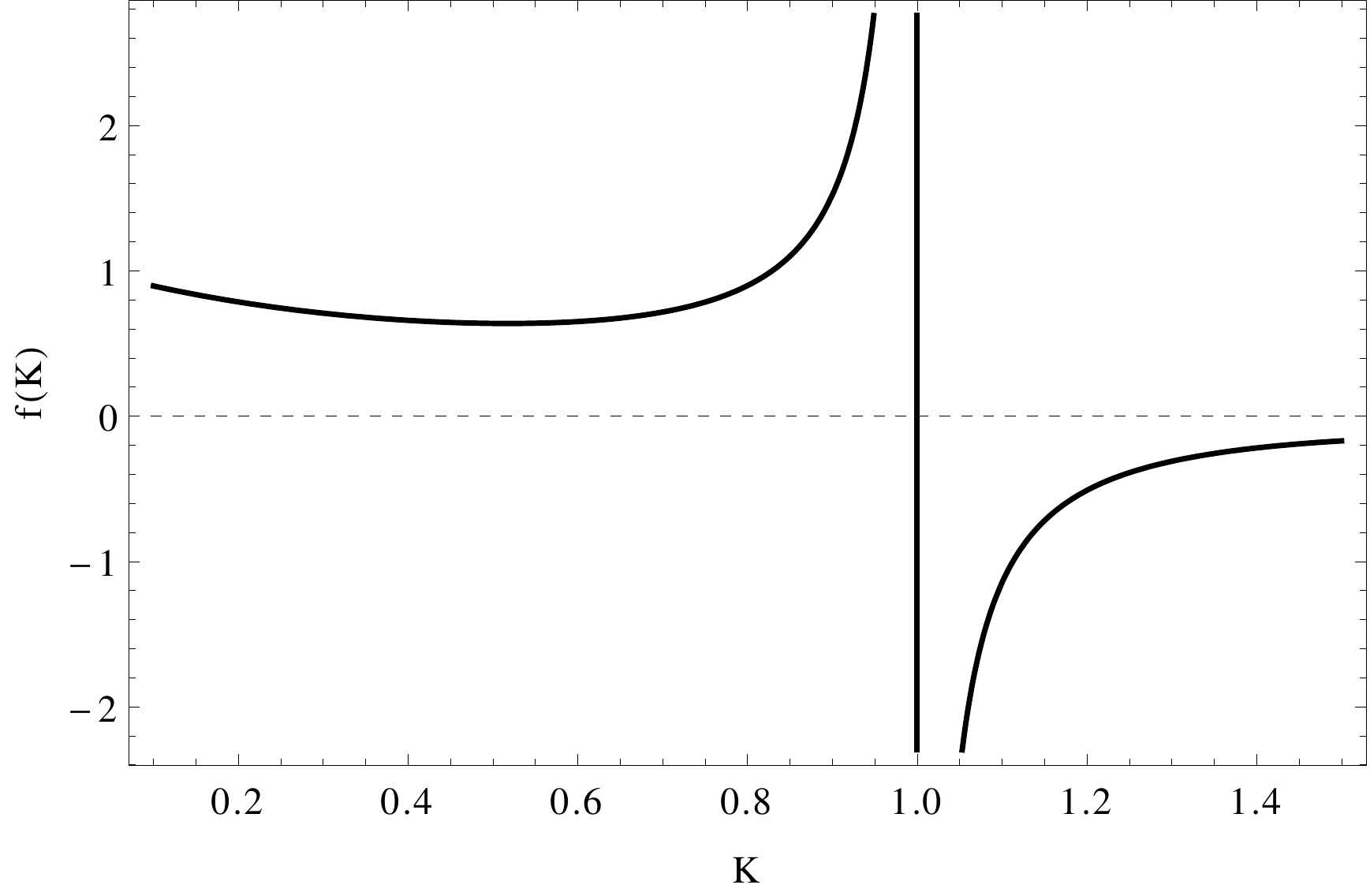}
      \caption{\small Function $f(K)$ describing the parametric dependence of conductivity of a clean HLL on the Luttinger liquid parameter K. 
      While $f(K)$ is singular at $K=1$ it always appears together with a function that vanishes at $K=1$ such that the expression for the conductivity 
      is finite in the noninteracting limit.
      \label{Fig:plot_f} }
\end{center}   
\end{figure}

In the case of a sufficiently clean sample the main mechanism of scattering will be $g_5$ umklapp interaction. Calculating the conductivity in the regime $\omega, T \ll k_F$ we obtain
\begin{align}
\begin{split} 
      \sigma(k \to 0, \omega) 
   =& \frac{i}{\omega^3} \frac{e^2 u^4 K^2}{h} \frac{2^6}{\pi^2}  \left( \frac{V}{u}\right)^2  
      \left( \frac{k_F}{k_0}\right)^2 \\
    & \times \frac{1}{(a k_0)^2}  \mathcal{I}_K(\omega,T) \label{refAppF}
\end{split}        
\end{align}
where the function $\mathcal{I}_K(\omega,T) $ is defined in Eq.~(\ref{IK}) in the Appendix. We can further simplify this result if the temperature is much higher or much lower than the frequency of the external field.
In the regime $\omega \gg T$ we obtain
\begin{align}
\begin{split} 
   \sigma(\omega) =& \frac{i}{\omega+i \delta} \frac{e^2 u}{h} \frac{8}{\pi^3} \left( \frac{V}{u}\right)^2 
                  \left( \frac{k_F}{k_0}\right)^2 (k_F a)^{2K} \\
                  & \times \frac{K^2}{(k_0 a)^2} (K-1) f(K) 
\end{split}                   
\end{align}
where we defined 
\begin{align}
\begin{split} 
   f(K) =& - \sin(K \pi) \bigl\lbrace \Gamma(-1-K)]^2 \\
         &+ (6+K)\Gamma(-K) \Gamma(-3-K) \bigr\rbrace .
\end{split}         
\end{align}
The function $f(K)$ is plotted in Fig.~\ref{Fig:plot_f}. While it is singular at $K=1$ it always appears together with a function that vanishes at $K=1$ such that the expression for the conductivity is finite in the noninteracting limit.

Since $\sigma$ is purely imaginary the only effect of the $g_5$ interaction is to renormalize the velocity $u$ and the Luttinger liquid parameter $K$ in the case without umklapp terms cf. Eq.~(\ref{sigma0}). 

In the regime $\omega \ll T$ we obtain
\begin{align}
\begin{split} 
   \Re \sigma(\omega) =& \frac{1}{\omega^2} \frac{e^2 u}{h} \frac{32}{\pi^3} \left( \frac{V}{u}\right)^2 
                  (k_F a)^{2K+2} \frac{u^2 k_F^2}{T} e^{-\frac{u k_F}{T}} \\
                  & \times \frac{K^2}{(k_0 a)^4}  \sin(K \pi) f(K) .
\end{split}                          
\end{align} 
In the limit of noninteracting electrons $K \to 1$ this agrees with the result obtained in the fermionic language Eq.~(\ref{conductivityinteraction1}) except for a nonuniversal constant of the order of unity.

Recall that the $g_5$ term in Eq.~(\ref{bosonicmodelclean}) is always irrelevant in the RG sense and therefore only yields small corrections to the fixed point properties. However, we observe that the nature of the corrections crucially depends on whether the RG flow is cut of by frequency or temperature if we are in the regime $\omega, T \ll k_F$. While frequency dependence only leads to a renormalization of the fixed point parameters $u$ and $K$, temperature dependence yields a finite conductivity which is however exponentially supressed. 

Finally, we can make some predictions about the regime of $\omega, T \gg k_F$. In this limit we can imply the result by the scaling dimension of $S_5$ in Eq.~(\ref{effectiveaction}) which yields $\sigma \sim \omega^{-2} (\max{(\omega,T}))^{2K+3}$. Comparing the limit $K \to 1$ with the fermionic case in Eq.~(\ref{conductivityinteraction2}) we see that this indeed gives the correct scaling of the conductivity and there is no cancellation in the leading order. Additionally, since temperature or frequency are much higher than the Fermi energy the system behaves effectively as if it were at the Dirac point, such that $k_F=0$. Therefore, we get
\begin{align}
   \sigma(\omega) \sim \frac{e^2 u}{h} \frac{1}{\omega^2}  u k_0 \left( \frac{V}{u}\right)^2 
                       \left(\frac{\max{(\omega,T})}{u k_0}\right)^{2K+3}.
\end{align}

\begin{figure}
\begin{center}
   \includegraphics[width=0.5\textwidth]{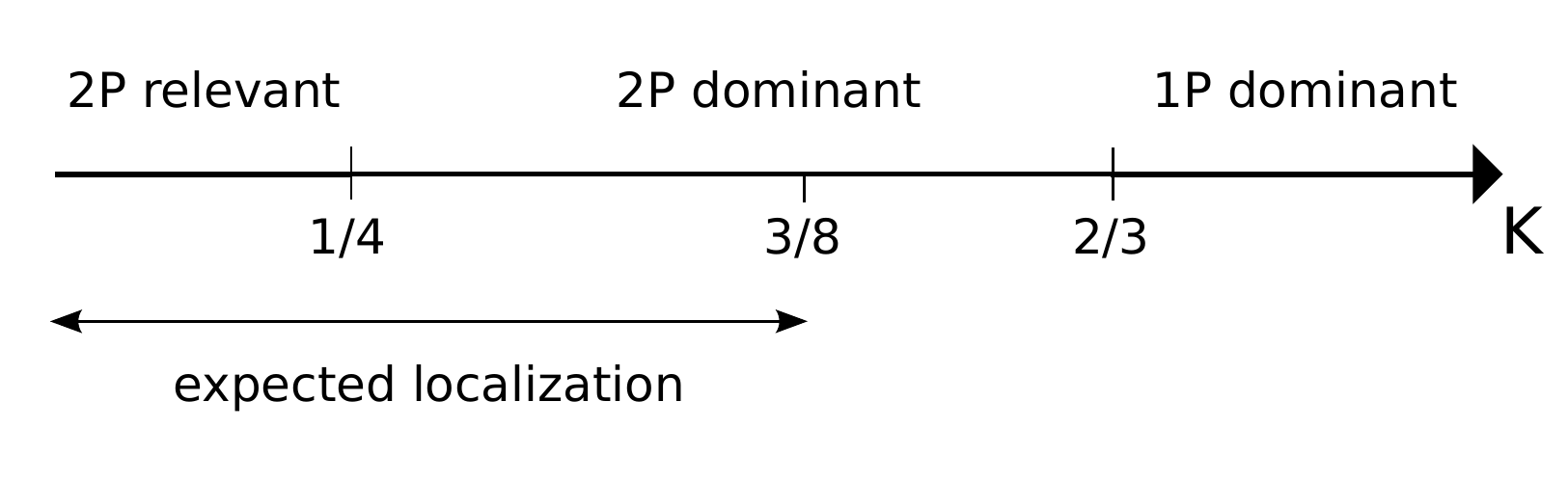}
      \caption{\small Phase diagram for the conductivity of a disordered helical liquid. 1P processes lead to a scattering time 
      $\tau_{\text{1P}} \sim T^{-2K-2}$ and 2P processes yield $\tau_{\text{2P}} \sim T^{-8K+2}$. At $K>2/3$ transport properties are dominated by 1P scattering.
      Below $K=2/3$ the 2P term has the lower scaling dimension and therefore becomes dominant. At $K=1/4$ the 2P term becomes relevant.
      As discussed in the main text a mapping to the Giamarchi-Schulz model of disordered LL suggests localization at K=3/8.  
      \label{Fig:phasediagram}      }
\end{center}   
\end{figure}

To summarize, we find that due to the strong correlation effects of the one dimensional Luttinger liquid the exponents of the power law in temperature now depend on the strength of interaction through the Luttinger liquid parameter K. In the limit of weakly interacting electrons $K \approx 1$ we reproduce the power law $T^5$ in Eq.~(\ref{conductivityinteraction2}) and the behavior in the limit $k_F \gg T$ in Eq.~(\ref{conductivityinteraction1}). Additionally, the present calculation in the bosonic form allows us to investigate the limit $\omega \gg T$. In the kinetic equation approach the external electric field is always treated classically, so it can not be applied if the corresponding frequency becomes larger than temperature. In this case one has to quantize the electric field and treat the interactions of photons with the system. While this treatment was not possible in the context of the kinetic equation, the quantum mechanical regime $\omega \gg T$ becomes accessible in the present Kubo formalism.

\subsection{{\it ac}  transport in the presence of disorder}
\label{sec:bosonization:subsec:disordered case}

In the presence of impurities we find the conductivity due to inelastic scattering processes in Appendix \ref{App:Calculation of the conductivity of a disordered helical Luttinger liquid}. The results read as
   \begin{align}
   \begin{split} 
         \sigma_{\text{2P}}(\omega,T) =& i \frac{e^2}{\omega^3}  32 u^2 K^2 g_{\text{2P}} 
      \left(\frac{\pi a T}{u} \right)^{8K} \mathcal{J}_{8K}(\omega,T)
   \end{split}   \label{referAppE1} \\
   \begin{split} 
         \sigma_{1P}(\omega,T) 
     =&  8 i \frac{e^2 u^2 K}{\pi a^4 } g_{\text{1P,1}} \frac{1}{\omega^3} \left(\frac{ \pi T}{u} \right)^{2K+4} \\
      &  \times  \left(3 \mathcal{J}_{2K+4}(\omega,T)- 2  \mathcal{J}_{2K+2}(\omega,T) \right) \label{referAppE2}
   \end{split}   
   \end{align}   
where $\mathcal{J}_{2K}(\omega,T)$ is defined in Eq.~(\ref{J2K}).

The limits of low and high temperature respectively are given by
\begin{widetext}
   \begin{align}
         \Re \sigma_{\text{2P}}(\omega,T) 
      =&  \frac{2 e^2 u }{h} \frac{1}{\omega^2}   \frac{4^3 K^4}{ \Gamma(8 K) \pi^4} 
         \left(\frac{V}{u}\right)^2 \frac{D_f}{u} \left(\frac{k_F}{k_0}\right)^2 \frac{1}{(a k_0)^6} 
        \begin{cases}
           \left(\frac{a \omega}{u} \right)^{8K-2}, &  \enspace \text{for} \enspace \omega \gg T \\ 
           \left(\frac{2 \pi a T}{u} \right)^{8K-2} \Gamma^2(4K), & \enspace \text{for} \enspace \omega \ll T
        \end{cases}, \\ \nonumber \\
         \Re \sigma_{\text{1P}}(\omega,T) 
   =&  \frac{2 e^2 u}{h} \frac{1}{\omega^2}  \left(\frac{2}{\pi}\right)^4 \frac{K^3}{\Gamma(2K+4)}  
       \left(\frac{V}{u}\right)^2  \frac{D_f}{u}\frac{1}{(a k_0)^4}       
   \begin{cases}
         3 \left(\frac{\omega a}{u}\right)^{2K+2},        &  \enspace \text{for} \enspace \omega \gg T  \\  
         \left(\frac{2 \pi a T}{u} \right)^{2K+2} K (K+1) \Gamma^2(K+1), & \enspace \text{for} \enspace \omega \ll T    
      \end{cases}   .
   \end{align}
\end{widetext}
Similarly to the clean case, we find power law exponents that depend on the strength of interactions through the Luttinger liquid parameter K.
Additionally, we observe that 2P scattering becomes the dominant scattering mechanism for $K<\frac{2}{3}$ and even becomes relevant for $K<\frac{1}{4}$. This behavior is in agreement with the results of Ref.~\onlinecite{lezmy12}. However, our derivation of these results from a more microscopic theory allows us to identify the origin of the 1P and 2P processes as the combined effect of $g_5$ and $g_3$ interaction together with scattering off impurities. In particular we identify the importance of forward scattering off disorder for transport properties, which has not been fully appreciated in the existing literature.

After having discussed the effect of interactions on transport, both by itself and in combination with forward scattering off disorder, we now comment on the effect of backscattering off the impurity described by the term $S_{\text{imp,b}}$ in Eq.~(\ref{bosonicmodel}).
In Appendix~\ref{App:Calculation of the conductivity of a disordered helical Luttinger liquid} we show that, to the leading order in disorder strength $D_b$, the backscattering term does not lead to a finite scattering time for any value of K. Recall that the term $S_{\text{imp,b}}$ originates from backscattering off impurities and should therefore have no impact on transport on its own, i.e. in the absence of $g_2$ interaction at K=1. However, we find that the conductivity does not only vanish for $ K = 1$ but for arbitrary $K$ meaning that even the combination of $g_2$ interaction and backscattering off impurities does not change the conductivity.
This is consistent with our fermionic analysis, see Table~\ref{Tab:results of second order perturbation theory }.

\subsection{Discussion of \textit{dc} conductivity and localization} 
\label{sec:bosonization:subsec:discussion}

In this section we complemented our previous kinetic equation calculation whose results are summarized in Tab.~\ref{Tab:summaryDC} by bosonizing the model for helical fermions and calculating the {\it ac} conductivity using the Kubo formula.
First, this allows us to treat Luttinger liquid renormalization effects that arise due to the strong correlations in one dimension and lead to power law exponents that depend on the strength of interaction through the Luttinger liquid parameter K. Second, it enables us to make predictions about the regime $\omega \gg T$ not captured by our previous kinetic equation analysis.

Before summarizing the results and discussing their implications for \textit{dc} transport we briefly comment on the effect of quantum interference phenomena on the transport properties of a disordered HLL. So far we have only discussed the quasiclassical regime where the dephasing length is much shorter than the mean free path.  Going beyond this semiclassical description we can also make predictions about localization in the helical Luttinger liquid. The model in the presence of disorder in Eq.~(\ref{bosonicmodel}) can be mapped onto the Giamarchi-Schulz model\cite{Giamarchi_Schulz_1988} of disordered LL with $K \to 4K$ by rescaling $\varphi$ fields. In combination with the analysis in Refs.~\onlinecite{Giamarchi_Schulz_1988,Gornyi_Polyakov_Mirlin_2007} this mapping suggests a transition to the localized state at $K= 3/8$.

Let us now summarize the perturbative results for the \textit{ac} conductivity and discuss their implications for \textit{dc} transport and the conductance of short edges channels.

In a sufficiently clean sample $g_5$ umklapp interaction leads to a power law behavior $\Re \sigma_{\textit{ac}} \sim \omega^{-2} \max{(\omega,T})^{2K+3}$ when the system is doped close to the Dirac point. In this case the kinetic equation treatment predicts Drude-like behavior of the conductivity and therefore we expect $\sigma_{\textit{dc}} \sim T^{-2K-3}$.

At $k_F \gg T$ i.e at filling far away from the Dirac point, we have to distinguish the regimes $\omega \gg T$ and $\omega \ll T$. If $\omega \gg T$, umklapp scattering does not lead to a finite real part of the conductivity. The only effect of the scattering process is then a renormalization of parameters $u$ and $K$.
On the other hand if $\omega \ll T$, the conductivity is exponentially suppressed and only the power law in front of the exponential is affected by Luttinger liquid renormalization. In either case we cannot make predictions about the {\it dc} conductivity since there exists an intermediary regime not captured by either approach.
 
In the presence of disorder we find the frequency and temperature dependence of the \textit{ac} conductivity as
\begin{align}
   \Re \sigma_{\textit{ac}} \sim \frac{1}{\omega^2} \begin{cases} \left[\max{(\omega,T)}\right]^{2K+2}, \, & \text{if} \enspace  K >2/3 ,\\ 
                                                                  \left[\max{(\omega,T)}\right]^{8K-2}, \, &\text{if} \enspace K<2/3 . \end{cases}
                                                                  \label{resultac}
\end{align}
Since the kinetic equation approach suggests that the parametric dependence of the conductivity is described by Drude's law, we predict the scaling of the semiclassical \textit{dc} conductivity as
\begin{align}
   \sigma_{\textit{dc}} \sim \begin{cases} \left[\max{(\omega,T)}\right]^{-2K-2}, \, & \text{if} \enspace  K >2/3 \\ 
                                           \left[\max{(\omega,T)}\right]^{-8K+2}, \, &\text{if} \enspace   K<2/3  \end{cases}
                                           \label{resultdc}
\end{align}

According to Eq.~(\ref{conductance}) we obtain the conductance of short edge channels from the \textit{ac} scattering time which yields
\begin{align}
   \delta G \sim \frac{e^2}{h} L  \enspace  \begin{cases} T^{-2K-2}, \, & \text{if} \enspace  K >2/3 \\ 
                                                T^{-8K+2}, \, &\text{if} \enspace K<2/3  \end{cases} \label{resultG}
\end{align} 
The complete phase diagram of the conductivity summarizing the transport properties in the presence of disorder is depicted in Fig.~\ref{Fig:phasediagram}.

\section{Summary and outlook} 
\label{sec:conclusion}
In this paper we have studied the transport properties of a generic one-dimensional helical liquid in the presence of interactions and disorder.

We have employed two complementing approaches for obtaining the conductivity in a wide range of parameters. One is a kinetic equation approach (Sec.~\ref{sec:Quantum kinetic equation formalism}, \ref{sec:Conductivity of helical fermions}) for weakly interacting helical fermions which allows us to determine the semiclassical conductivity in the regime $\omega \ll T$ both in the \textit{ac} and \textit{dc} limit. The results of this treatment are summarized in Tab.~\ref{Tab:summaryDC}. The other approach is bosonization (Sec.~\ref{sec:LL}, \ref{sec: Results bosonization}) combined with the linear response Kubo formalism which enables us to include Luttinger liquid renormalization effects as well as to describe the regime $\omega \gg T$, where the external electric field cannot be treated classically anymore. 
By combining the two approaches we have demonstrated that while the helical liquid is topologically protected against elastic scattering events, inelastic scattering that arises due to the combined effect of interactions and disorder leads to a finite conductivity.

In a clean helical Luttinger liquid, we find that $g_5$ interaction leads to a finite conductivity. Due to a peculiar kinetics necessarily involving a particle at the Dirac point, the parametric dependence of conductivity induced by this term cannot be described by Drude's law. This is discussed in detail in Sec.~\ref{subsec: Comparison and discussion} and we include Luttinger liquid effects in Sec.~\ref{sec:bosonization:subsec:clean case}. 

Our main result is the phase diagram for the conductivity of a disordered HLL depicted in Fig.\ref{Fig:phasediagram} and the corresponding temperature or frequency dependence in Eqs.~(\ref{resultac}) and~(\ref{resultdc}). We find that the parametric dependence of the conductivity of a disordered HLL as a function of frequency is described by Drude's law where the temperature or frequency dependence is a power law with exponents depending on the Luttinger liquid parameter K. This behavior arises due to combined effects of interaction and impurity scattering. 
Thereby, it is of conceptual importance that forward scattering off disorder, in contrast to disorder induced backscattering, plays the primary role in these combined effects. An intuitive physical explanation for this fact is yet to be formulated.
 
During our analysis we assumed a weak-disorder limit, $D_b,D_f \ll k_F$, and studied the theory in the leading order in $D_b$ and $D_f$. We expect that the effect of higher-order terms amounts to a renormalization of the couplings in the effective field theory; a detailed study is left for future work.

Going beyond the semiclassical regime, we make predictions about localization in a one-dimensional helical liquid by employing a mapping to the Giamarchi-Schulz model of disordered Luttinger liquid. This suggests a localization transition at $K=3/8$.
A detailed analysis of localization in helical edge states remains a prospect for future work.

\section{Acknowlegdgements} 
\label{sec:Acknowlegdgements}
We thank S. Rachel, B. Trauzettel, D. Aristov, F. Crepin, F. Geissler, M. Schuett and D. Polyakov for useful and interesting discussions. NK acknowledges the hospitality of the Institute for Theoretical Physics at Wuerzburg and financial support by the Carl-Zeiss-Stiftung. This work was supported by the program DFG SPP 1666 ``Topological insulators", by the German-Israeli Foundation and by BMBF. When this work was completed we became aware of the preprint,\cite{Geissler_Crepin_Trauzettel_2014} where the effect of Rashba impurities on transport properties was discussed. We thank B. Trauzettel for sharing the results of Ref.~\onlinecite{Geissler_Crepin_Trauzettel_2014} prior to its publication.

\appendix 
\begin{widetext}

\section{Kinetic equation: Calculation of {\it ac} conductivity}
\label{App:AC conductivity due to Umklapp interaction}
In this Appendix we demonstrate how to obtain the \textit{ac} conductivity for weakly interacting electrons in the context of a kinetic equation.

First, we summarize the symmetry properties of some objects relevant for subsequent calculations
   \begin{itemize}
      \item The Fermi-Dirac distribution obeys $f^{(0)}_{k\eta} = f^{(0)}_{-k\bar{\eta}}$.
      \item In the absence of scattering i.e. when the collision integral vanishes, the symmetries of a
            solution $\psi$ of Eq.~(\ref{KE2}) are determined by the driving term 
            $eE \eta f^0_{\eta,k} (1-f^0_{\eta,k})/T$ and therefore:
         \begin{align}
            \psi_{k,\eta} = - \psi_{-k,\bar{\eta}} \label{symmetrypsi}.
         \end{align}
             We checked explicitly that there exist solutions with this symmetry even in the presence of 
             relaxation inducing processes. In the following calculations we will only consider
             solutions that obey the above symmetry.
      \item The object $\Gamma_{12,1'2'}$ defined in Eq.~(\ref{Gamma}) is invariant under exchange of the first and second two arguments
            e.g. $\Gamma_{12,1'2'} = \Gamma_{21,1'2'}$, which is obvious from its definition.
            Under the assumption of (\ref{symmetrypsi}) it is straightforward to show that 
            $\Gamma_{12,1'2'} = - \Gamma_{-1,-2,-1',-2'}$ where $-1 \equiv (-k_1,\bar{\eta}_1)$. 
   \end{itemize} 
To calculate the {\it ac} conductivity we proceed as follows. First we calculate the transition matrix element $\mathcal{M}_{1,2,1',2'} = \bra{1' 2'} T \ket{12}$ between initial and final momentum eigenstates due to the scattering processes in the T-matrix. From this we obtain the collision integral using Eq.~(\ref{collisionintegral2P}) and finally the conductivity using Eq.~(\ref{sigmaAC}).
In the following we use the notation $1 \equiv (k_1,\eta_1)$ and define $\ket{0}$ as the groundstate of the free hamiltonian.    
We define the following integrals often encountered during the calculation of {\it ac} conductivity
\begin{align}
   \lambda_1 =& \int \! \mathrm{d} x_1  \mathrm{d} x_2 \mathrm{d} x_3 \, (x_1-x_2)^2 n_F(x_1) n_F(x_2) \left[1-n_F(-x_3)\right] \left[1-n(x_1+x_2+x_3) \right]
               \simeq 103.9, \label{lambda1}\\
   \lambda_2 =& \int \! \mathrm{d} x_1  \mathrm{d} x_2 \mathrm{d} x_3 \, (x_1-x_2)^2 (x_1+x_2+ 2 x_3)^2  n_F(x_1) n_F(x_2) \left[1-n_F(-x_3)\right] 
                \left[1-n(x_1+x_2+x_3) \right]
               \simeq 1757.97    . \label{lambda2}  
\end{align}

In the first order of the expansion in the T-matrix, Eq.~(\ref{Tmatrix}), we consider only interaction setting $T = H_{\text{int}}$ and show explicitily the calculation of \textit{ac} conductivity for $T= H_3$ and $T= H_5$.
\subsection{$g_3$ term}

For $g_3$ interaction we obtain the matrix element
\begin{align}
       \prescript{(a)}{}{\mathcal{M}_{1,2,1',2'}} =&  \bra{1' 2'} H_3 \ket{12}   
   =   \frac{V}{k_0^4 L} \sum_{k,p,q,\eta} \left(k^2-(k-q)^2 \right) \left(p^2-(p+q)^2 \right)
       \prescript{(a)}{}{\mathcal{M}_{1,2,1',2'}^{k,p,q,\eta}},    
\end{align}
where we defined
\begin{align}
\begin{split} 
     \prescript{(a)}{}{\mathcal{M}_{1,2,1',2'}^{k,p,q,\eta}}
   =& \bra{0} \psi_{1'}^{} \psi_{2'}^{} \psi_{k,\eta}^{\dagger} \psi_{p,\eta}^{\dagger} \psi_{p+q,\bar{\eta}}^{}
             \psi_{k-q,\bar{\eta}}^{} \psi_{1}^{\dagger} \psi_{2}^{\dagger} \ket{0} \\
   =& \delta_{\eta,\eta_{1'}} \delta_{\eta,\eta_{2'}} \delta_{\bar{\eta},\eta_{1}} \delta_{\bar{\eta},\eta_{2}}
     \left( \delta_{k,k_{2'}} \delta_{p,k_{1'}} - (1' \leftrightarrow 2') \right) 
     \left( \delta_{k-q,k_{1}} \delta_{p+q,k_{2}} - (1 \leftrightarrow 2) \right).        
\end{split}     
\end{align}
Using this result we obtain
\begin{align}
       \prescript{(a)}{}{\mathcal{M}_{1,2,1',2'}} 
   =   \frac{V}{k_0^4 L} \sum_{\eta} \delta_{\eta,\eta_{1'}} \delta_{\eta,\eta_{2'}} \delta_{\bar{\eta},\eta_{1}}
       \delta_{\bar{\eta},\eta_{2}} \delta_{k_1+k_2, k_{1'}+ k_{2'}} h_{k_1,k_2,k_{1'},k_{2'}} ,      
\end{align}
where we defined $ h_{k_1,k_2,k_{1'},k_{2'}} = 2 (k_1-k_2) (k_{1'}^2-k_{2'}^2) (k_1+k_2-2(k_{1'}+ k_{2'}))  $.     
The corresponding collision integral is
\begin{align}
   I_1[\psi] = -2 \pi \frac{V^2}{k_0^8 L^2} \sum_{k_2,k_{1'},k_{2'}} \delta_{k_1+k_2, k_{1'}+ k_{2'}}
               h^2_{k_1,k_2,k_{1'},k_{2'}} \Gamma_{(k_1,\eta_1),(k_2,\eta_1),(k_{1'},\bar{\eta}_1),(k_{2'},\bar{\eta}_1)}.
\end{align}
To obtain the conductivity we have to calculate the object
\begin{align}
\begin{split}
     &  \frac{1}{L} \sum_{k_1,\eta_1} \eta_1 I_1[\psi^{(0)}] \\
    =& \, 4 \pi \frac{V^2}{k_0^8}  \frac{ e E}{(-i \omega) T}
       \int \! \frac{\mathrm{d} k_1}{2 \pi} \frac{\mathrm{d} k_2}{2 \pi}  \frac{\mathrm{d} k_{1'}}{2 \pi} \,
        h^2_{k_1,k_2,k_{1'},k_1+k_2 -k_{1'}} f^{(0)}_{k_1,R} f^{(0)}_{k_2,R} 
       (1-f^{(0)}_{k_{1'},L} ) (1-f^{(0)}_{k_1+k_2-k_{1'},L} ) \delta( k_1 + k_2) = 0.       
\end{split}       
\end{align}
In the last equality we used that $h(k_1,-k_1,k_{1'},-k_{1'}) = 0$. Consequently, $g_3$ interaction alone does not affect transport.

\subsection{$g_5$ term}
\label{App:g5fermionic}

In order to calculate the transition matrix element for $g_5$ we need
\begin{align}
\begin{split} 
   \prescript{(b1)}{}{\mathcal{M}_{1,2,1',2'}^{k,p,q,\eta}}
   =& \bra{0} \psi_{1'}^{} \psi_{2'}^{} \psi_{k+q,\eta}^{\dagger} \psi_{p-q,\bar{\eta}}^{\dagger} \psi_{p,\eta}^{}
             \psi_{k,\eta}^{} \psi_{1}^{\dagger} \psi_{2}^{\dagger} \ket{0} \\
   =& \delta_{\eta,\eta_{1}} \delta_{\eta,\eta_{2}} 
     \left( \delta_{\eta,\eta_{2'}} \delta_{\bar{\eta},\eta_{1'}}  \delta_{k+q,k_{2'}} \delta_{p-q,k_{1'}} 
     - (1' \leftrightarrow 2') \right) 
     \left( \delta_{k,k_{1}} \delta_{p,k_{2}} - (1 \leftrightarrow 2) \right)  , 
\end{split}     \\
\begin{split} 
   \prescript{(b2)}{}{\mathcal{M}_{1,2,1',2'}^{k,p,q,\eta}}
   =& \bra{0} \psi_{1'}^{} \psi_{2'}^{} \psi_{k,\eta}^{\dagger} \psi_{p,\eta}^{\dagger} \psi_{p-q,\bar{\eta}}^{}
             \psi_{k+q,\eta}^{} \psi_{1}^{\dagger} \psi_{2}^{\dagger} \ket{0} \\
   =& \delta_{\eta,\eta_{1'}} \delta_{\eta,\eta_{2'}} 
     \left( \delta_{\eta,\eta_{1}} \delta_{\bar{\eta},\eta_{2}}  \delta_{k+q,k_{1}} \delta_{p-q,k_{2}} 
     - (1 \leftrightarrow 2) \right) 
     \left( \delta_{k,k_{2'}} \delta_{p,k_{1'}} - (1' \leftrightarrow 2') \right)  .       
\end{split}           
\end{align}
The matrix element is then given by
\begin{align}
\begin{split} 
       \prescript{(b)}{}{\mathcal{M}_{1,2,1',2'}} 
   =&  -\frac{V}{k_0^2 L} \sum_{k,p,q,\eta}  \eta \left(k^2-p^2 \right) \left[ \prescript{(b1)}{}
       {\mathcal{M}_{1,2,1',2'}^{k,p,q,\eta}} + \prescript{(b2)}{}{\mathcal{M}_{1,2,1',2'}^{k,p,q,\eta}} \right] \\
   =&  -\frac{2 V}{k_0^2 L} \sum_{\eta} \eta \, \delta_{k_1+k_2, k_{1'}+ k_{2'}} 
      \Bigl[ (k_1^2-k_2^2) \delta_{\eta,\eta_1} \delta_{\eta,\eta_2} 
      \left( \delta_{\bar{\eta},\eta_{1'}} \delta_{\eta,\eta_{2'}} -\delta_{\bar{\eta},\eta_{1'}} \delta_{\eta,\eta_{2'}}
       \right) \\
    &  + (k_{2'}^2-k_{1'}^2) \delta_{\eta,\eta_{1'}} \delta_{\eta,\eta_{2'}} 
      \left( \delta_{\bar{\eta},\eta_{2}} \delta_{\eta,\eta_{1}} -\delta_{\bar{\eta},\eta_{1}} \delta_{\eta,\eta_{2}}
       \right) \Bigr].
\end{split}       
\end{align}
From this we obtain the collision integral as
\begin{align}
\begin{split}
   I_1[\psi] =& -2 \pi \frac{4 V^2}{k_0^4 L^2} \sum_{k_2,k_{1'},k_{2'}} \delta_{k_1+k_2, k_{1'}+ k_{2'}}
               \Bigl\lbrace (k_1^2-k_2^2)^2 
                  \left[  \Gamma_{(k_1,\eta_1),(k_2,\eta_1),(k_{1'},\bar{\eta}_1),(k_{2'},\eta_1)} 
                        + \Gamma_{(k_1,\eta_1),(k_2,\eta_1),(k_{1'},\eta_1),(k_{2'},\bar{\eta}_1)}  \right] \\
              & \hspace{4.7cm}             +(k_{1'}^2-k_{2'}^2)^2 
                  \left[  \Gamma_{(k_1,\eta_1),(k_2,\bar{\eta}_1),(k_{1'},\eta_1),(k_{2'},\eta_1)} 
                        + \Gamma_{(k_1,\eta_1),(k_2,\bar{\eta}_1),(k_{1'},\bar{\eta}_1),(k_{2'},\bar{\eta}_1)}  \right]
               \Bigr\rbrace.
\end{split}               
\end{align}
Notice that from the 10 total terms in the absolute square of the matrix element only those terms with the same chirality Kronecker deltas survive the summation over external chiralities.

In order to obtain conductivity we have to calculate the quantity  $\sum_{k_1,\eta_1} \eta_1 I_1[\psi^{(0)}]$. 
Using the symmetry arguments for $\Gamma$ defined at the beginning of the Appendix and the abbreviation $\left\lbrace k \right\rbrace = k_1,k_2,k_3,k_4$ we find
\begin{align}
\begin{split} 
    &   \sum_{\left\lbrace k \right\rbrace} (k_1^2-k_2^2) \Bigl( \Gamma_{(k_1,R),(k_2,R),(k_{1'},L),(k_{2'},R)}
      + \Gamma_{(k_1,R),(k_2,R),(k_{1'},R),(k_{2'},L)} - \Gamma_{(k_1,L),(k_2,L),(k_{1'},R),(k_{2'},L)} \\
      &- \Gamma_{(k_1,L),(k_2,L),(k_{1'},L),(k_{2'},R)} \Bigr) 
   = 4 \sum_{\left\lbrace k \right\rbrace} (k_1^2-k_2^2) \Gamma_{(k_1,R),(k_2,R),(k_{1'},L),(k_{2'},R)},
\end{split}    \\
\begin{split} 
    &   \sum_{\left\lbrace k \right\rbrace} (k_{1'}^2-k_{2'}^2) \Bigl( \Gamma_{(k_1,R),(k_2,L),(k_{1'},R),(k_{2'},R)}
      + \Gamma_{(k_1,R),(k_2,L),(k_{1'},L),(k_{2'},L)} - \Gamma_{(k_1,L),(k_2,R),(k_{1'},L),(k_{2'},L)} \\
    &  - \Gamma_{(k_1,L),(k_2,R),(k_{1'},R),(k_{2'},R)} \Bigr)
   = 0 .
\end{split}   
\end{align}
With this we can simplify the expression yielding
\begin{align}
\begin{split} 
        \frac{1}{L} \sum_{k_1,\eta_1} \eta_1 I_1[\psi^{(0)}]
   =&   \frac{32 \pi V^2}{k_0^4} \frac{ eE}{(-i \omega ) T} 
        \int \! \frac{\mathrm{d} k_1}{2 \pi}  \frac{\mathrm{d} k_{2}}{2 \pi} 
        \frac{\mathrm{d} k_{1'}}{2 \pi}\,   (k_1^2 -k_2^2)^2 f^{(0)}_{k_1,R} f^{(0)}_{k_2,R} (1-f^{(0)}_{k_{1'},L})
        (1-f^{(0)}_{k_1+k_2-k_{1'},R})  \delta(k_{1'}) \\
   =& \frac{V^2 e E}{k_0^4 (-i \omega)  h} T^5 f(\zeta), 
\end{split}   
\end{align}
where $f(\zeta)$ is defined in the main text in Eq.~(\ref{f(zeta)}). The real part of the conductivity is then given as
\begin{align}
\begin{split}
 \Re \sigma_{\text{{\it ac}}} =& \frac{e}{E L  \omega} \Im \sum_{k, \eta} \eta I_{k, \eta}[\psi^{(0)}]\\
                        =& \frac{e^2 v_F }{h} \frac{1}{\omega^2}  \left(\frac{V}{v_F} \right)^2
                           v_F k_0 \left(\frac{T}{v_F k_0} \right)^5 f(\zeta) ,  
\end{split}                           
\end{align}
where we reinstated $v_F$ in the last line. This result is used in Eq.~(\ref{referAppA}) of the main text.

\subsection{$g_5$ interaction combined with forward scattering}
\label{App:Discussion $g_5$ with forward scattering}
In the second order of the T-matrix expansion in Eq.~(\ref{Tmatrix}) we we have to include the following transition matrix  elements
\begin{align}
     \bra{1'2'} H_{\text{int}} G_0 H_{\text{int}} \ket{12} + \bra{1'2'} H_{\text{imp}} G_0 H_{\text{int}} \ket{12}+
     \bra{1'2'} H_{\text{int}} G_0 H_{\text{imp}} \ket{12} + \bra{1'2'} H_{\text{imp}} G_0 H_{\text{imp}} \ket{12}  
\end{align} 
Since the system we consider is time reversal symmetric processes containing only disorder will not affect transport properties, so will will not consider the term $\bra{1'2'} H_{\text{imp}} G_0 H_{\text{imp}} \ket{12}$. Additionally, we will neglect the term $\bra{1'2'} H_{\text{int}} G_0 H_{\text{int}} \ket{12}$ containing only disorder, since we already obtained results for the conductivity in the first order expansion of the T-matrix and the second order will be subleading in interaction strength $V$.

Therefore, we are left with terms containing both scattering due to interaction and disorder. Thereby, $H_{\text{imp}}$ contains forward and backward scattering off disorder and $H_{\text{int}}$ contains all g-ology terms defined in Eq.~(\ref{model}) and we have to consider arbitrary combinations of the two. We remark that only combined processes that change the chirality of at least one incoming particle lead to a finite conductivity. The results for the conductivity induced by these combined processes is summarized in Table \ref{Tab:results of second order perturbation theory }. 

In this Appendix we choose $H_{\text{imp}} = H_f$ and $H_{\text{int}} = H_5$ as an example to demonstrate the calculations performed to obtain the \textit{ac} conductivity due to combined processes. 

We start by defining effective states containing disorder as follows
\begin{align}
   \ket{12}_f =& G_0 H_{\text{imp,f}} \ket{12} = \frac{U}{L} \sum_q \left\lbrace \frac{1}{\bar{\eta}_1 q +i \delta} 
                 \psi_{k_1+q,\eta_1}^{\dagger} \psi_{2}^{\dagger} -(1 \leftrightarrow 2) \right\rbrace \ket{0}, \\
   \ket{12}_b =& G_0 H_{\text{imp,b}} \ket{12} = \frac{2 k_F}{k_0^2}\frac{U}{L} \sum_q \left\lbrace 
                 \frac{2 k_1+q}{\eta_1 (2 k_1+q) +i \delta} \psi_{k_1+q,\eta_1}^{\dagger} \psi_{2}^{\dagger} 
                 -(1 \leftrightarrow 2) \right\rbrace \ket{0}.            
\end{align}
Furthermore, we consider momenta close to the Fermi surface and simplify the $g_5$ term as
\begin{align}
   H_5 = \frac{2 V k_F }{k_0^2 L} \sum_{k,p,q} \sum_{\eta} (p-k) \psi_{k+q,\eta}^{\dagger} \psi_{p-q,\bar{\eta}}^{\dagger}
         \psi_{p,\eta}^{} \psi_{k,\eta}^{} +h.c. \, \equiv \tilde{H}_5 +\tilde{H}_5^{\dagger}.  \label{refg5}  
\end{align} 
When considering the combination of $g_5$ and forward scattering we have to add the following transition matrix elements
\begin{align}
     \bra{1' 2'} \tilde{H}_5 \ket{12}_f + \bra{1' 2'} \tilde{H}_5^{\dagger} \ket{12}_f 
   + \prescript{}{f}{\bra{1' 2'}} \tilde{H}_5 \ket{12} + \prescript{}{f}{\bra{1' 2'}} \tilde{H}_5^{\dagger} \ket{12} .
\end{align}
Notice that the matrix elements are connected through complex conjugation as 
\begin{align}
     \left[ \bra{1' 2'} \tilde{H}_5 \ket{12}_f \right]^{\dagger} \equiv \prescript{(1)}{}{\mathcal{R}_{1,2,1',2'}}^{\ast}
   = \prescript{}{f}{\bra{1 2}} \tilde{H}_5^{\dagger} \ket{1'2'} \equiv \prescript{(1)}{}{\mathcal{L}_{1',2',1,2}}^{\ast}  
     \enspace \text{with} \enspace \delta \to -\delta .
\end{align}
In the last line we took into account that complex conjugation changes the retarded to an advanced Greens function i.e.
\begin{align}
   \ket{12}_f = G_0^{(R)} H_{\text{imp,f}} \ket{12} \stackrel{\ast}{\to} \bra{21} H_{\text{imp,f}} G_0^{(A)} =
   \prescript{}{f}{\bra{21}}.
\end{align}
Here, $\delta$ denotes the infinitesimal self energy of the free retarded Greens function.

Consequently we only calculate the matrix elements $\mathcal{R}$ where the effective ket is to the right. The other matrix elements are obtained by exchanging $1 \leftrightarrow 1'$, $2 \leftrightarrow 2'$ and $\delta \to -\delta$.
We obtain
\begin{align}
\begin{split}
      \prescript{(1)}{}{\mathcal{R}_{1,2,1',2'}^{k,p,q,q',\eta} }
   =& \bra{0} \psi_{k_{1'},\eta_{1'}}^{} \psi_{k_{2'},\eta_{2'}}^{} \psi_{k+q,\eta}^{\dagger} 
             \psi_{p-q,\bar{\eta}}^{\dagger} \psi_{p,\eta}^{} \psi_{k,\eta}^{}
             \psi_{k_1+q',\eta_1}^{\dagger} \psi_{k_2,\eta_2}^{\dagger} \ket{0} - (1 \leftrightarrow 2)\\
   =& \delta_{\eta_1,\eta} \delta_{\eta_2,\eta}\Bigl( \delta_{\eta_{1'},\bar{\eta}} \delta_{\eta_{2'},\eta}
      \delta_{k+q,k_{2'}} \delta_{p-q,k_{1'}} -(1' \leftrightarrow 2') \Bigr)
      \Bigl( \delta_{k_1+q',k} \delta_{p,k_2}- \delta_{k_1+q',p} \delta_{k,k_2} \Bigr)  - (1 \leftrightarrow 2),
\end{split}      \\
\begin{split} 
      \prescript{(2)}{}{\mathcal{R}_{1,2,1',2'}^{k,p,q,q',\eta} }
   =& \bra{0} \psi_{k_{1'},\eta_{1'}}^{} \psi_{k_{2'},\eta_{2'}}^{} \psi_{k,\eta}^{\dagger} 
             \psi_{p,\eta}^{\dagger} \psi_{p-q,\bar{\eta}}^{} \psi_{k+q,\eta}^{}
             \psi_{k_1+q',\eta_1}^{\dagger} \psi_{k_2,\eta_2}^{\dagger} \ket{0} - (1 \leftrightarrow 2)\\
   =& \delta_{\eta_{1'},\eta} \delta_{\eta_{2'},\eta}\Bigl( \delta_{\eta_{1},\eta} \delta_{\eta_{2},\bar{\eta}}
      \delta_{k+q,k_{1}+q'} \delta_{k_2,p-q} -\delta_{\eta_{2},\eta} \delta_{\eta_{1},\bar{\eta}}
      \delta_{k+q,k_{2}} \delta_{k_1+q',p-q} \Bigr)
      \Bigl( \delta_{k_{1'},p} \delta_{k_{2'},k}- (1' \leftrightarrow 2') \Bigr) \\
    &   - (1 \leftrightarrow 2).   
\end{split}      
\end{align}
The corresponding transition matrix element reads as
\begin{align}
\begin{split} 
       \mathcal{M}^{\mathcal{R}}_{1,2,1',2'} 
   =&  \frac{2 k_F V U}{k_0^2 L^2} \sum_{k,p,q,q'} \sum_{\eta} \left\lbrace \frac{p-k}{\bar{\eta}_1 q' + i \delta}
       \left(\prescript{(1)}{}{\mathcal{R}_{1,2,1',2'}^{k,p,q,q',\eta}} +
            \prescript{(2)}{}{\mathcal{R}_{1,2,1',2'}^{k,p,q,q',\eta}}   \right)
       \right\rbrace \\
   =&  \frac{2 k_F V U}{k_0^2 L^2} \sum_{\eta} 
   \Bigl[
       \frac{1}{\bar{\eta}_1 (k_{1'}+k_{2'} -k_1-k_2) + i\delta} 
       \Bigl\lbrace 4(k_2-k_1) \delta_{\eta_1,\eta} \delta_{\eta_2,\eta} (\delta_{\eta_{1'},\bar{\eta}} 
       \delta_{\eta_{2'},\eta} -\delta_{\eta_{1'},\eta} \delta_{\eta_{2'},\bar{\eta}}) \\
    &  +2(k_{2'}-k_{1'}) \delta_{\eta_{1'},\eta} \delta_{\eta_{2'},\eta} (\delta_{\eta_{1},\bar{\eta}} 
       \delta_{\eta_{2},\eta} -\delta_{\eta_{1},\eta} \delta_{\eta_{2},\bar{\eta}})  \Bigr\rbrace \\
    &  -\frac{2 (k_{2'} -k_{1'})}{\bar{\eta}_2 (k_{1'}+k_{2'} -k_1-k_2) + i\delta}
       \delta_{\eta_{1'},\eta} \delta_{\eta_{2'},\eta} (\delta_{\eta_{2},\bar{\eta}} 
       \delta_{\eta_{1},\eta} -\delta_{\eta_{2},\eta} \delta_{\eta_{1},\bar{\eta}})   
   \Bigr]      
\end{split}                             
\end{align}
As discussed we imply $\mathcal{M}^{\mathcal{L}}_{1,2,1',2'}$ by setting $1 \leftrightarrow 1'$, $2 \leftrightarrow 2'$ and $\delta \to -\delta$ in the above result. Now, we use the identity $2 \Re \frac{1}{x+i\delta} = \frac{x}{x^2+\delta^2} \equiv \mathcal{P} \frac{1}{x}$ to obtain
\begin{align}
\begin{split} 
   \mathcal{M}_{1,2,1',2'} =& \mathcal{M}^{\mathcal{R}}_{1,2,1',2'}  +\mathcal{M}^{\mathcal{L}}_{1,2,1',2'} \\
                           =& \frac{2 k_F V U}{k_0^2 L^2} \sum_{\eta} 
                              \mathcal{P}\frac{1}{\bar{\eta}_1 (k_{1'}+k_{2'} -k_1-k_2)}  
                              \Bigl[ (k_2-k_1)\delta_{\eta_1,\eta} \delta_{\eta_2,\eta} (\delta_{\eta_{1'},\bar{\eta}} 
                                 \delta_{\eta_{2'},\eta} -\delta_{\eta_{1'},\eta} \delta_{\eta_{2'},\bar{\eta}}) \\
                            &  +(k_{2'}-k_{1'}) \delta_{\eta_{1'},\eta} \delta_{\eta_{2'},\eta} 
                              (\delta_{\eta_{1},\bar{\eta}} \delta_{\eta_{2},\eta} -\delta_{\eta_{1},\eta}
                               \delta_{\eta_{2},\bar{\eta}})   \Bigr]   . 
\end{split}                               
\end{align}
The collision integral reads as
\begin{align}
\begin{split} 
   I_1[\psi^{(0)}] =& -2 \pi N_{\text{imp}} \sum_{1',2',2} \Gamma_{1,2,1',2'} \left|\mathcal{M}_{1,2,1',2'}\right|^2 \\
                   =& -2 \pi N_{\text{imp}} \left(\frac{2 k_F U V}{k_0^2 L^2} \right)^2 \sum_{k_{1'},k_{2'},k_2} \left(
                       \mathcal{P}\frac{1}{\bar{\eta}_1 (k_{1'}+k_{2'} -k_1-k_2)}  \right)^2 \\
                    &  \Bigl\lbrace (k_1-k_2)^2 (\Gamma_{(k_1,\eta_1),(k_2,\eta_1),(k_{1'},\bar{\eta}_1),(k_2',\eta_1)}
                       +\Gamma_{(k_1,\eta_1),(k_2,\eta_1),(k_{1'},\eta_1),(k_2',\bar{\eta}_1)}) \\
                    &  +(k_{1'}-k_{2'})^2 (\Gamma_{(k_1,\eta_1),(k_2,\bar{\eta}_1),(k_{1'},\bar{\eta}_1),
                      (k_2',\bar{\eta}_1)} +\Gamma_{(k_1,\eta_1),(k_2,\bar{\eta}_1),(k_{1'},\eta_1),(k_2',\eta_1)})
                      \Bigr\rbrace .   
\end{split}                                      
\end{align}
Using the symmetry properties of $\Gamma$ we obtain
\begin{align}
   \sum_{k_1,\eta_1} \eta_1 I_1[\psi^{(0)}]= 
   \frac{2}{\pi^2} \frac{e E}{(-i \omega) T h} n_{\text{imp}} \left(\frac{2 k_F U V}{k_0^2 } \right)^2 T^3 g(\zeta) ,
\end{align}
where
\begin{align}
   g(\zeta) = \int_{-\zeta}^{\zeta} \! \mathrm{d} x_1 \mathrm{d} x_2 \mathrm{d} x_3 \, 
              \frac{4 x_3^2}{(4 x_3^2+\delta^2)^2} (x_1-x_2)^2 n_F(x_1-\zeta) n_F(x_2-\zeta) (1-n_F(-x_3-\zeta)) 
              (1-n_F(x_1+x_2+x_3-\zeta)) .
\end{align}
First we calculate the integral over $x_3$ by sending the integration limits to infinity and completing the contour in the complex plane. We find
\begin{align}
   g(x_1,x_2,\zeta) = \frac{i \pi}{2} n_B(-x_1-x_2+2 \zeta) \sum_{n=-\infty}^{\infty} \Bigl\lbrace
                      \frac{1}{(i \pi (2n+1)-\zeta)^2}-\frac{1}{(i \pi (2n+1)-x_1-x_2+\zeta)^2}
                      \Bigr\rbrace +\mathcal{O}\left(\frac{1}{\delta}\right), 
\end{align}
where we defined $n_B(x) =\frac{1}{e^x-1}$. The expression is formally divergent when sending $\delta$ to zero. However, this divergency will be regularized by taking into account a finite electronic self energy, due to impurity scattering or interactions. Thus we will neglect the $1/\delta$ part in the following.

We proceed by using the series representation of the polygamma function $\psi^{(n)}(z) = (-1)^{n+1} n! \sum_{k=0}^{\infty} \frac{1}{(z+k)^{n+1}}, \enspace n>0$ to rewrite the expression as
\begin{align}
   g(x_1,x_2,\zeta) = -\frac{i }{4 \pi} n_B(-x_1-x_2+2 \zeta) \Bigl\lbrace
                      \psi^{(1)}\left(\frac{1}{2}-\frac{\zeta}{2 \pi i}  \right) -
                      \psi^{(1)}\left(\frac{1}{2}-\frac{x_1+x_2-\zeta)}{2 \pi i}\right)
                      \Bigr\rbrace. 
\end{align}
Using the asymptotics of the first polygamma function $\psi^{(1)}(z) \sim z^{-1}$ for $|z| \gg 1$ and the fact that $x_1,x_2 \ll \zeta$ we obtain
\begin{align}
   g(\zeta) \approx - \frac{1}{\zeta^2} \frac{1}{2} \int_{-\infty}^{\infty} \!  \mathrm{d} x_1 \mathrm{d} x_2 \,
                      (x_1-x_2)^2 (x_1+x_2) n_F(x_1) n_F(x_2) n_B(-x_1-x_2) 
            \approx 51.9 \zeta^{-2}  .        
\end{align}
The resulting conductivity is therefore 
\begin{align}
   \Re \sigma(\omega,T) = \frac{e }{E \omega} n_{\text{imp}} \Im \sum_{1} \eta_1 I_1[\psi^{(0)}]
                    = 42.1 \frac{e^2 v_F}{h} \frac{1}{\omega^2}  n_{\text{imp}} \left(\frac{U V}{v_F^2} \right)^2
                      \left( \frac{T}{v_F k_0} \right)^4.
\end{align}
This result constitutes Eq.~(\ref{conductivitycombined1}) of the main text.
\section{Kinetic equation: Calculation of \textit{dc} conductivity}
\label{App:Kinetic equation: Calculation of dc conductivity}
The purpose of this Appendix is to calculate the \textit{dc} conductivity of a weakly interacting helical liquid by using exact integral equations for the fermionic distribution function obtained from the solution of a kinetic equation.

First, we calculate the transition matrix element $\mathcal{M}_{1,2,1',2'} = \bra{1' 2'} T \ket{12}$, for $T=H_5$, $T=H_{\text{1P}}$ and $T= H_{\text{2P}}$. The corresponding terms in the Hamiltonian are defined in Eq.~(\ref{refg5}), Eq.~(\ref{H1P}) and Eq.~(\ref{H2P}). We then obtain the collision integral using Eq.~(\ref{collisionintegral2P}).
The results read as

\begin{align}
   \begin{split} 
       I_1^{(g_5)}[\psi] 
   =& -8 \pi \frac{ V^2}{k_0^4 L^2} \sum_{k_2,k_{1'},k_{2'}} \delta_{k_1+k_2, k_{1'}+ k_{2'}}
       \Bigl\lbrace 2 (k_1^2-k_2^2)^2 
                   \Gamma_{(k_1,\eta_1),(k_2,\eta_1),(k_{1'},\bar{\eta}_1),(k_{2'},\eta_1)}     \\
    &              +(k_{1'}^2-k_{2'}^2)^2
                   \left[  \Gamma_{(k_1,\eta_1),(k_2,\bar{\eta}_1),(k_{1'},\eta_1),(k_{2'},\eta_1)} 
                   + \Gamma_{(k_1,\eta_1),(k_2,\bar{\eta}_1),(k_{1'},\bar{\eta}_1),(k_{2'},\bar{\eta}_1)}  \right]
       \Bigr\rbrace, \label{CIg5} 
    \end{split}   \\
    \begin{split} 
      I_1^{(1P)}[\psi]
   =& -4 \pi n_{\text{imp}} (U V)^2 \frac{1}{k_0^4 L^3} \sum_{k_2,k_{1'},k_{2'}} 
               \Bigl\lbrace 2 (k_1-k_2)^2 
                    \Gamma_{(k_1,\eta_1),(k_2,\eta_1),(k_{1'},\bar{\eta}_1),(k_{2'},\eta_1)}     \\
    &            +(k_{1'}-k_{2'})^2 \left[  \Gamma_{(k_1,\eta_1),(k_2,\bar{\eta}_1),(k_{1'},\eta_1),(k_{2'},\eta_1)} 
                 + \Gamma_{(k_1,\eta_1),(k_2,\bar{\eta}_1),(k_{1'},\bar{\eta}_1),(k_{2'},\bar{\eta}_1)}  \right]
               \Bigr\rbrace ,\label{CI1P}
   \end{split}      \\
     I_1^{(2P)}[\psi]
   =& -128 \pi n_{\text{imp}} (U V)^2 \frac{k_F^2}{k_0^8 L^3} \sum_{k_2,k_{1'},k_{2'}} (k_{2'} - k_{1'})^2 (k_2-k_1)^2
      \Gamma_{(k_1,\eta_1),(k_2,\eta_1),(k_{1'},\bar{\eta}_1),(k_{2'},\bar{\eta}_1)} .  \label{CI2P}  
\end{align}
Here, $\Gamma$ is defined in Eq.~(\ref{Gamma}) and contains the thermal factors for the specific process, the distribution function $\psi$ and the energy conserving delta function.
\subsection{Clean case: $g_5$ interaction}
\label{App:Kinetic equation: Calculation of dc conductivity:clean case}
Let us first consider a clean system where only $g_5$ influences transport. We insert Eq.~(\ref{CIg5}) into Eq.~(\ref{psi}) to get an integral equation for $\psi_{1}$.

\begin{align}
\begin{split}
   \psi_{k_1} = -4 \frac{V^2}{k_0^4} \frac{1}{(-i \omega)} \frac{1}{f_{k_1,R}^{(0)}(1-f_{k_1,R}^{(0)}) } \Big\lbrace
            &\int \! \frac{\mathrm{d} k_2}{2 \pi} \, \mathcal{K}(k_1,k_2) \left[ \psi_{k_1} + \psi_{k_2} + \psi_{0} -\psi_{k_1+k_2} \right]  \\
           +&\int \! \frac{\mathrm{d} k_1'}{2 \pi} \, \mathcal{L}(k_1,k_{1'}) \left[ \psi_{k_1} - \psi_{0} - \psi_{k_{1'}} -\psi_{k_1-k_{1'}} \right]  \\
           + &\delta(k_1) \int \! \frac{\mathrm{d} k_2}{2 \pi} \, \mathcal{C}(k_2,k_{1'}) \left[ \psi_{0} - \psi_{-k_2} + \psi_{-k_{1'}} +\psi_{k_{1'}-k_2}
            \right]  \Big\rbrace \label{psig5}
\end{split}            
\end{align}
Here, we defined
\begin{align}
\mathcal{K}(k_1,k_2) =& (k_1^2-k_2^2)^2 f^{(0)}_{k_1,R} f^{(0)}_{k_2,R} (1-f^{(0)}_{0,L}) (1-f^{(0)}_{k_1+k_2,R}),\\
\mathcal{L}(k_1,k_{1'}) =& \frac{1}{2} (k_{1'}^2-(k_1-k_{1'})^2)^2 f^{(0)}_{k_1,R} f^{(0)}_{0,L} (1-f^{(0)}_{k_{1'},R}) 
                                 (1-f^{(0)}_{k_1-k_{1'},R}),\\
\mathcal{C}(k_2,k_{1'}) =& \frac{1}{2} (k_{1'}^2-(k_{1'}-k_2)^2)^2 f^{(0)}_{0,R} f^{(0)}_{k_2,L} (1-f^{(0)}_{k_{1'},L}) (1-f^{(0)}_{k_2-k_{1'},L}).
\end{align}
Due to the delta function $\delta(k_1)$ in the third line of Eq.~(\ref{psig5}) we have to treat the distribution function $\psi_{k_1=0}$ of the state at the Dirac point separately.

First, we consider states at $k_1 \neq 0$ and introduce dimensionless momenta  $x_i = k_i /T$ and $\zeta = k_F/T$ which yields
\begin{align}
\begin{split}
   \tilde{\psi}_{\zeta}(x_1) =& -\frac{\mathcal{A}_-(x_1,\zeta)}{\mathcal{A}_{+}(x_1,\zeta)} \tilde{\psi}_{\zeta}(0) 
                             - \frac{1}{\mathcal{A}_{+}(x_1,\zeta)} \int \! \frac{\mathrm{d} x_2}{2 \pi} 
                                \, \mathcal{B}(x_1,x_2,\zeta) \tilde{\psi}_{\zeta}(x_2)
                             -\frac{n_F(x_1-\zeta) (1-n_F(x_1-\zeta))}{\mathcal{A}_{+}(x_1,\zeta)} .
                            \label{IEg51}
\end{split}
\end{align}
where we defined 

\begin{align}
\begin{split} 
   \tilde{\psi}(x)        =&  4 V^2 \frac{1}{k_0^4} \frac{1}{e E} T^{6} \psi(x),    \\
   \mathcal{A}_{\pm}(x_1,\zeta)   =& \int \! \frac{\mathrm{d} x_2}{2 \pi} \, 
                                 \left[ \mathcal{L}(x_1,x_2,\zeta) \pm \mathcal{K}(x_1,x_2,\zeta) \right],\\
   \mathcal{B}(x_1,x_2,\zeta) =& \mathcal{K}(x_1,x_2,\zeta)-\mathcal{K}(x_1,x_2-x_1,\zeta) 
                                 -2 \mathcal{L}(x_1,x_2,\zeta).   
\end{split}                                 
\end{align}
We observe that the zero momentum distribution function $\psi_{\zeta}(0)$ explicitly affects the distribution function of all other momentum states.
In order to obtain $\psi_{\zeta}(0)$ we have to consider the case of zero external momentum, $k_1=0$, in Eq.~(\ref{psig5}).

In this case we have to regularize the diverging delta function. Physically, the divergence stems from our assumption of an infinite system where momentum is a continuous variable. We therefore introduce a momentum cutoff $\Lambda$ such that $\delta(x=0)= \Lambda $ and neglect the other contributions in the integral equation for $\psi_{\zeta}(0)$ in comparison to this term. 
Solving the resulting equation for $\psi_{\zeta}(0)$ yields the cutoff independent result, 
\begin{align}
   \tilde{\psi}_{\zeta}(0) = -\frac{1}{\mathcal{D}(\zeta)} \int \! \frac{\mathrm{d} x_2}{2 \pi} \,
                            \mathcal{E}(x_2,\zeta) \tilde{\psi}_{\zeta}(x_2) \label{IEg52},
\end{align}
where
\begin{align}
\begin{split} 
   \mathcal{D}(\zeta)         =& \int \! \frac{\mathrm{d} x_1 \mathrm{d} x_2}{2 \pi} \, \mathcal{C}(x_1,x_2,\zeta),\\
   \mathcal{E}(x_1,\zeta)     =& \int \! \mathrm{d} x_2 \, \left\lbrace -\mathcal{C}(-x_1,x_2,\zeta) 
                                 +2 \mathcal{C}(x_2,-x_1,\zeta) \right\rbrace.
\end{split}                                 
\end{align}

We now insert the result for the zero momentum distribution function in Eq.~(\ref{IEg52}) into the integral equation determining the distribution function of the remaining states, Eq.~(\ref{IEg51}). This yields
\begin{align}
\begin{split} 
   \tilde{\psi}_{\zeta}(x_1) =& \frac{\mathcal{A}_-(x_1,\zeta)}{\mathcal{A}_{+}(x_1,\zeta)} \frac{1}{\mathcal{D}(\zeta)} 
                              \int \! \frac{\mathrm{d} x_2}{2 \pi} \, \mathcal{E}(x_2,\zeta) 
                              \tilde{\psi}_{\zeta}(x_2) 
                             - \frac{1}{\mathcal{A}_{+}(x_1,\zeta)} \int \! \frac{\mathrm{d} x_2}{2 \pi} \,
                                \mathcal{B}(x_1,x_2,\zeta) \tilde{\psi}_{\zeta}(x_2) \\
                             &-\frac{n_F(x_1-\zeta) (1-n_F(x_1-\zeta))}{\mathcal{A}_{+}(x_1,\zeta)} .
\end{split}
\end{align}
This is an exact integral equation determining the distribution function of helical fermions in the presence of $g_5$ interaction. While it can not be solved analytically we can solve it numerically in the regime of high and low temperatures yielding a solution $\tilde{\psi}_{\zeta}$.

In terms of this dimensionless function $\tilde{\psi}_{\zeta}$ the conductivity in Eq.~(\ref{sigmaDC}) takes the form
\begin{align}
   \sigma_{\text{\it dc}} =& - \frac{2 e^2}{h} \frac{k_0^4}{T^5} \frac{1}{4 V^2} \kappa(\zeta) \\
   \kappa(\zeta) =& \int \! \mathrm{d} x \, \, n_F(x-\zeta) (1-n_F(x-\zeta)) \tilde{\psi}_{\zeta}(x)
\end{align}
We numerically find the asymptotics
\begin{align}
   \kappa(\zeta) \simeq \begin{cases} \, -3.23 \enspace \zeta^{-5} e^{\zeta}, \enspace &\zeta \gg 1 \\
                                      \, -0.056,                     \enspace &\zeta = 0  \end{cases} 
\end{align}
As an example of the quality of the obtained asymptotics we plot the quantity $ \tilde{\kappa}(\zeta) = \zeta^5 e^{-\zeta} \kappa(\zeta)$ which converges to $\tilde{\kappa}(\zeta \to \infty)= -3.23$ in Fig.~(\ref{Fig:g5dc}).

The obtained limits of $\kappa(\zeta)$ yield the expression for conductivity in the regime $k_F \ll T$, 
\begin{align}
   \sigma(k_F \ll T) = 0.014 \times  \frac{2 e^2 v_F}{h} \left( \frac{v_F}{V} \right)^2  \frac{1}{v_F k_0}
                   \left( \frac{v_F k_0}{T} \right)^5,
\end{align}
and in the regime $k_F \gg T$,
\begin{align}
   \sigma(k_F \gg T) = 0.81 \times \frac{2 e^2 v_F}{h} \left( \frac{v_F}{V} \right)^2 
                       \left(\frac{k_0}{k_F}\right)^{4} \frac{1}{v_F k_F} e^{\frac{v_F k_F}{T}}.
\end{align} 
These results are used in the main text in Eq.~(\ref{referAppA1}) and Eq.~(\ref{referAppA2}).
\begin{figure}
\begin{center}
   \includegraphics[width=0.45\textwidth]{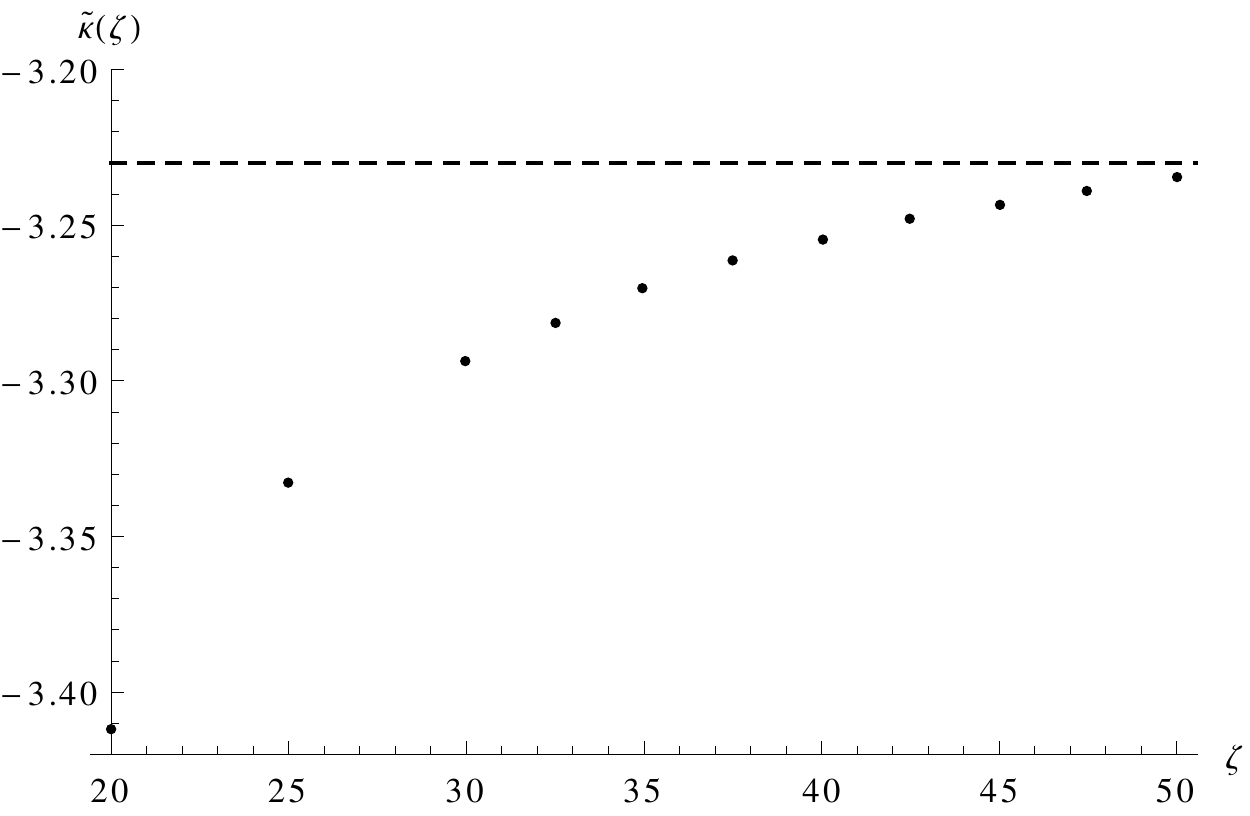}
      \caption{\small The asymptotics of the function $\tilde{\kappa}(\zeta)$, defined in the main text in the context of the {\it dc} conductivity of a 
                      clean system, as a function of the ratio of Fermi energy and temperature, $\zeta = k_F/T$. We observe that it converges to a value 
                      $\kappa_4(\zeta \to \infty) = -3.23$. 
      \label{Fig:g5dc}      }
\end{center}   
\end{figure}

\subsection{Disordered case: effective 1P and 2P processes}
\label{App:Kinetic equation: Calculation of dc conductivity:disordered case}

Inserting the collision integrals for inelastic single and two particle processes defined in in Eq.~(\ref{CI1P}) and (\ref{CI2P}) into Eq.~(\ref{psi}) we obtain integral equations describing the distribution function $\psi$.

After introducing dimensionless quantities and taking the limit $\omega \to 0$ the equations take the form
   \begin{align}
     \begin{split}  \tilde{\psi}_i(x_1+\zeta) 
    =& -\frac{1}{\mathcal{H}_i(x_1)}  \int \! \mathrm{d} x_2 \, \mathcal{G}_i(x_1,x_2) \tilde{\psi}_i(x_2+\zeta) 
      -\frac{1}{\mathcal{H}_i(x_1)}  n_F(x_1) (1-n_F(x_1)) \label{IEDCdisorder} .\end{split}
   \end{align}
Here, $i=1,2$ denotes 1P and 2P processes, respectively and we have defined
\begin{align}
\begin{split} 
     \tilde{\psi}_i(x)        =&  \frac{\bar{g}_{\text{i}}^2}{e E} T^{3+2 i} \psi_i(x), \\
     \mathcal{F}(x_1,x_2,x_3) =&  n_F(x_1) n_F(x_2) (1-n_F(-x_3)) (1-n_F(x_1+x_2+x_3)) 
                                  \left[(x_1-x_2)^2  (2x_3+x_1+x_2)^2 \right], \\
     \mathcal{G}_1(x_1,x_2)   =&  \int \! \frac{\mathrm{d} x_3}{2 \pi} \, n_F(x_1) n_F(x_2) (1-n_F(-x_3))
                                  (1-n_F(x_1+x_2+x_3)) \left[(x_1-x_2)^2 - (2x_3+x_1+x_2)^2 \right], \\     
     \mathcal{G}_2(x_1,x_2)   =&  \int \! \frac{\mathrm{d} x_3}{2 \pi} \, \left( \mathcal{F}(x_1,x_3,x_2)+2 
                                  \mathcal{F}(x_1,-x_2,x_3)\right), \\
     \mathcal{H}_1(x_1)       =&  \int \! \frac{\mathrm{d} x_3 \mathrm{d} x_2}{(2 \pi)^2} \,  n_F(x_1) n_F(x_2) 
                                  (1-n_F(-x_3)) (1-n_F(x_1+x_2+x_3)) \left[(x_1-x_2)^2 + (2x_3+x_1+x_2)^2 \right], \\
     \mathcal{H}_2(x_1)       =&  \int \! \frac{\mathrm{d} x_3 \mathrm{d} x_2}{(2 \pi)^2} \, \mathcal{E}(x_1,x_3,x_2).
\end{split}     
\end{align} 
In the presence of disorder the physics of the scattering processes is not sensitive to the ratio of Fermi energy and temperature and thus the functions $\tilde{\psi}_i$ are in fact $\zeta$ independent. 

The {\it dc} conductivity due to 1P ($i=1$) or 2P ($i=2$) processes is then obtained as
   \begin{align}
      \sigma_i =& -\frac{2 e^2}{ h} \frac{1}{\bar{g}_i^2 T^{2i+2}} \kappa_i, \\
      \kappa_i =&  \int \! \mathrm{d} x \, n_F(x) (1-n_F(x)) \tilde{\psi}_i(x+\zeta) \label{kappai}
   \end{align}
where $\kappa_1 = -0.46$ and  $\kappa_2 = -0.042$. To calculate the $\kappa_i$ we first solved the integral equations, Eq.~(\ref{IEDCdisorder}) numerically and subsequently used the obtained solutions $\tilde{\psi}_i$ to get $\kappa_i$ according to Eq.~(\ref{kappai}).

This procedure yields the {\it dc} conductivity in the presence of impurities:
\begin{align}
   \sigma_{\text{1P}} =& \frac{\kappa_1 }{2} \times \frac{ 2 e^2 v_F}{h} \frac{1}{v_F n_{\text{imp}}}  
                         \left(\frac{v_F^2}{ U V} \right)^2 \left(\frac{v_F k_0}{T} \right)^4, \\
   \sigma_{\text{2P}} =& \frac{\kappa_2}{2^6} \times  \frac{2 e^2 v_F}{h} \frac{1}{v_F n_{\text{imp}}}  
                         \left(\frac{v_F^2}{ U V} \right)^2 \left( \frac{k_0}{k_F}\right)^2
                         \left(\frac{v_F k_0}{T} \right)^6.                     
\end{align}
These results constitute Eq.~(\ref{referAppA3}) and Eq.~(\ref{referAppA4}) of the main text.

\section{Average over forward scattering in the bosonic action}
\label{App:Average over forward scattering in the bosonic action}
In this Appendix we perform the disorder average over forward scattering in the effective low energy action of a disordered helical liquid in Eq.~(\ref{effectiveaction}).
After gauging out forward scattering according to Eq.~(\ref{gaugeoutforward}) and averaging over backward scattering the action reads as
   \begin{align}
   \begin{split}
      S_3 =& \frac{4 k_F^2 V}{\pi^2 a^4 k_0^4 }   \sum_a \int \! \mathrm{d} x \mathrm{d} \tau \,
             \cos(2 \sqrt{4 \pi} \varphi_a +\frac{4K}{u} \int_{x_0}^x \! \mathrm{d} y \, U_f(y) -4k_F x) ,\\
      S_5 =& \frac{ 4 V k_F }{ \pi^{\frac{3}{2}} a k_0^2 } \sum_a \int \! \mathrm{d} x \mathrm{d} \tau \,
             \partial_x^2 \theta_a \sin(\sqrt{4 \pi} \varphi_a + \frac{2 K}{u }\int_{x_0}^x \! \mathrm{d} y \, U_f(y) 
             -  2 k_F x) \\
             -& \frac{ 16 V k_F^2 }{ \pi^{\frac{3}{2}} a k_0^2 } \sum_a  \int \! \mathrm{d} x \mathrm{d} \tau \, 
              \partial_x \theta_a  \cos(\sqrt{4 \pi} \varphi_a+ \frac{2 K}{u }\int_{x_0}^x \! \mathrm{d} y \, U_f(y)
             -  2 k_F x) ,\\
      S_{b} =& -\frac{4 D k_F^2}{\pi a^2 k_0^4} \sum_{a,b} \int \! \mathrm{d} x \mathrm{d} \tau \mathrm{d} \tau' \,
                \partial_x \theta_a(x,\tau) \partial_x \theta_b(x,\tau')
                \cos(\sqrt{4 \pi} (\varphi_a(x,\tau) - \varphi_b(x,\tau'))).      
    \end{split}              
   \end{align}

At this point we still have to average over forward scattering. To investigate the relevant averages consider the toy action
\begin{align}
    S =& \, g \int \! \mathrm{d} x \mathrm{d} \tau \,  ( e^{i \int^x U} + e^{-i \int^x U}), \\
    \overline{U(x) U(x')} =& \,  \delta(x-x').
\end{align}
We perform the disorder average perturbatively in g: 
\begin{equation}
   \overline{e^{-S} }\approx 1-\overline{S}+\frac{1}{2} \overline{S^2} + \cdots \approx e^{-\overline{S}+\frac{1}{2} \left(\overline{S^2}-\overline{S}^2
   \right)}.
\end{equation}
Now $\overline{S^n}$ contains terms $\overline{e^{i  \sum_{m=0}^n \alpha_m \int^{x_m} \! \mathrm{d} y \, U(y)   } }$ where $\alpha_m \in \left\lbrace 1,-1 \right\rbrace$. Using the auxiliary identity, 
\begin{equation}
   \int_{x_0}^{x} \! \mathrm{d} y  \int_{x_0}^{x'} \! \mathrm{d} y' \, \delta(y-y') = \min(x,x')-x_0,
\end{equation}
we find
\begin{align}
      \overline{e^{i \alpha_m \sum_{m=0}^n \int^{x_m} \! \mathrm{d} y \, U(y)   } } 
   =  e^{-\frac{1}{2}  \left( \sum_{m=0}^n \sum_{l=0}^n \alpha_m \alpha_l  
       \int^{x_m} \! \mathrm{d} y \, \int^{x_l} \! \mathrm{d} y' \, \overline{U(y) U(y')} \right)   } 
   =  e^{-\frac{1}{2} \left( \sum_{m,l} \alpha_m \alpha_l (\min(x_m,x_l) -x_0) \right)}.
   \end{align}
In the limit $x_0 \to -\infty$ this is only nonzero if $\sum_{m,l}^n \alpha_m \alpha_l = 0$ i.e.
\begin{align}
   \sum_{m,l}^n \alpha_m \alpha_l = \sum_{m=l} \alpha_m^2 + \sum_{m \neq l} \alpha_m \alpha_l 
                                  = n+ 2 \sum_{m < l} \alpha_m \alpha_l \stackrel{!}{=} 0.
\end{align}
Notice that the second term is even, i.e.  $\overline{S^n}$ vanishes for odd n. For the lowest nontrivial order we find
\begin{align}
   \overline{S^2} =  e^{-\frac{1}{2} \overline{\left( \alpha \int^x U +\alpha' \int^{x'} U 
                                   \right)^2}} 
                  =  e^{-\frac{1}{2} (x+x'-2\min(x,x'))} \delta_{\alpha,-\alpha'}
                  =   e^{-\frac{1}{2} |x-x'|} \delta_{\alpha,-\alpha'}.
\end{align}
To summarize:\\
Let us define $\tilde{U}_{\alpha}(x) = e^{i \alpha \int_{x_0}^x \! \mathrm{d} y \, U_f(y) } $ where $\alpha \in \mathbb{R}$. We showed that $\tilde{U}$ obeys gaussian statistics up to fourth order in a weak coupling expansion.
 \begin{align}
      \begin{split} \overline{\tilde{U}_{\alpha}(x) \tilde{U}_{\alpha}(x')} =& 0,\\
      \overline{\tilde{U}_{\alpha}(x) \tilde{U}^{\ast}_{\alpha}(x')} =& e^{-\frac{\alpha^2}{2} D_f |x-x'|}  .\end{split}
   \end{align}
Thus nonlocal interactions decay exponentially due to forward scattering off disorder.
The resulting interaction terms in the action are of the form
   \begin{align}
      S =& -g \int \! \mathrm{d} x \mathrm{d} \tau \, \int \! \mathrm{d} x' \mathrm{d} \tau' \, 
                        F[\varphi(x,\tau),\varphi(x',\tau')]  e^{-D_f \mu |x-x'|} e^{-i \nu k_F (x-x')} , 
   \end{align}
where F is some functional of the fields and $\mu, \nu$ are constants. Now we split the spatial integration into relative and center of mass coordinates $R = \frac{1}{2} (x+x')$, $r= x-x'$. The relevant scales for the low energy physics of the model are given by energies much smaller than the disorder strength $D_f$. That means we can assume that the fields in F are smooth as a function of the relative coordinate r, 
\begin{align}
\begin{split}
    S =&  -g \int \! \mathrm{d} r \mathrm{d} R  \mathrm{d} \tau \mathrm{d} \tau' \,   F[\varphi(r,R,\tau),\varphi(r,R,\tau')] 
         e^{-D_f \mu |r|} e^{-i \nu k_F r} \\
      \approx&  -g \int \! \mathrm{d} R  \mathrm{d} \tau \mathrm{d} \tau' \, F[\varphi(R,\tau),\varphi(R,\tau')]            
       \left( \int_{-\infty}^{\infty} \! \mathrm{d} r \,e^{-D_f \mu |r|} 
         e^{-i \nu k_F r} \right)\\
      =& -\frac{g}{\pi^2 } \frac{\mu }{\nu^2 } \frac{D_f}{k_F^2} \int \! \mathrm{d} R  \mathrm{d} \tau
          \mathrm{d} \tau' \, F[\varphi(R,\tau),\varphi(R,\tau')]   .    
\end{split}          
\end{align}
This procedure yields the local theory discussed in Eq.~(\ref{bosonicmodel}), where our new momentum cutoff is given by the strength of forward scattering off disorder $D_f$.

\section{Formalism for conductivity calculation in the bosonized language}
\label{App:Formalism for conductivity calculation in the bosonized language}
In this Appendix we state some general methods and formulas needed to calculate the {\it ac} conductivity in the bosonized language.

\subsection{Anomalous current and susceptibility}

In order to compute the anomalous contributions to the current and the diamagnetic susceptibility we perform the minimal substitution $\partial_x \theta \to \partial_x \theta + \frac{e}{\sqrt{\pi}} A$ in the model for a clean HLL, Eq.~(\ref{bosonicmodelclean}) and in the model describing the disordered HLL, Eq.~(\ref{bosonicmodel}). The current $j$ and diamagnetic susceptibility $\chi^{\text{dia}}$ are then obtained by varying with respect to the vector potential, $j = \delta S/\delta A|_{A=0}$ and $\chi^{\text{dia}}(x-x',\tau-\tau') = - \frac{\delta S}{\delta A(x,\tau) \delta A(x',\tau')}$. This yields
\begin{align}
    j_{\text{an,clean}}(1) =& -\frac{4 e V k_F}{\pi^2 a k_0^2} \partial_{x_1} \sin(\sqrt{4 \pi} \varphi(1)-  2 k_F x_1)
                                      -\frac{ 16 e V k_F^2 }{ \pi^2 a k_0^2 } \partial_{x_1} \theta(1)  
                                       \cos(\sqrt{4 \pi} \varphi(1)-  2 k_F x_1)  , \\
    \chi^{\text{dia}}_{\text{an,clean}}(1-2) =& \,
                                      \frac{ 16 e^2 V k_F^2 }{ \pi^{\frac{5}{2}} a k_0^2 } \cos(\sqrt{4 \pi} \varphi(1)-  2 k_F x_1)
                                      \delta(1-2)                                        
\end{align}
in the clean case and
   \begin{align}
              \begin{split} \left[j_{\text{an,dis}}\right]_a(x_1,\tau_1)       
            =& \, 2 g_{\text{1P,1}} \frac{e}{\sqrt{\pi}} \sum_{b} \int \! \mathrm{d} \tau \, \partial_{x_1} 
              \left(  \partial^2_{x_1} \theta_b(x_1,\tau) \cos\left\lbrace \sqrt{4 \pi} \left[ \varphi_a(x_1,\tau_1) 
              - \varphi_b(x_1,\tau)\right]\right\rbrace \right) \\
              &+2 g_{\text{1P,2}} \frac{e}{\sqrt{\pi}} \sum_{b} \int \! \mathrm{d} \tau \, \partial^2_{x_1} \theta_b(x_1,\tau) 
              \sin\left\lbrace \sqrt{4 \pi} \left[ \varphi_a(x_1,\tau) - \varphi_b(x_1,\tau_1)\right]\right\rbrace  \\
              &+2 g_{\text{1P,2}} \frac{e}{\sqrt{\pi}} \sum_{b} \int \! \mathrm{d} \tau \, \partial_{x_1} \theta_b(x_1,\tau) \partial_{x_1} 
              \sin\left\lbrace \sqrt{4 \pi} \left[ \varphi_a(x_1,\tau) - \varphi_b(x_1,\tau_1)\right]\right\rbrace  \\
            & -2 g_{\text{imp,b}} \frac{e}{\sqrt{\pi}} \sum_{b} \int \! \mathrm{d} \tau \, 
                \partial_{x_1} \theta_b(x_1,\tau) \cos \left\lbrace \sqrt{4 \pi} 
                \left[ \varphi_a(x_1,\tau_1) - \varphi_b(x_1,\tau) \right] \right\rbrace   \end{split}\\
              \begin{split} \left[\chi^{\text{dia}}_{\text{an,dis}}\right]_{ab}(1-2)       
            =& \, 2 g_{\text{1P,2}} \frac{e^2}{\pi} \delta(x_1-x_2) \, \partial_{x_1} 
               \sin\left\lbrace \sqrt{4 \pi} \left[ \varphi_a(x_1,\tau_2) - \varphi_b(x_1,\tau_1)\right]\right\rbrace  \\
            & +2 g_{\text{imp,b}} \frac{e^2}{\pi} \cos \left\lbrace \sqrt{4 \pi} 
                \left[ \varphi_a(x_1,\tau_1) - \varphi_b(x_1,\tau_2) \right] \right\rbrace  \delta(x_1-x_2)  \end{split}    
   \end{align}
in the disordered case. Here, we abbreviated $1 = (x_1,\tau_1)$. These expressions are needed to obtain the \textit{ac} conductivity in Appendix~\ref{App:Calculation of the conductivity of a disordered helical Luttinger liquid} and \ref{App:Calculation of the conductivity of a clean helical Luttinger liquid}.

\subsection{Correlation functions}

In order to calculate the correlation functions that appear during the calculation of conductivity we state some basic correlation functions of the bosonic theory, which can be obtained using standard methods.\cite{Giamarchi_Book, Delft_Schoeller_Review} 
\begin{align}
\begin{split} 
  \braket{\partial_x \varphi(x,\tau) \partial_x \varphi(0,0)}
  =&  -\frac{K}{4 \pi} \left( \frac{\pi T}{u} \right)^2 \left(\frac{1}{\sinh^2(x_+)} + \frac{1}{\sinh^2(x_-)} \right),\\
   \braket{\partial^2_x \varphi(x,\tau) \partial^2_x \varphi(0,0)}
  =&  \, \frac{K}{2 \pi} \left( \frac{\pi T}{u} \right)^4 \left(\frac{1+ 2\cosh^2(x_+)}{\sinh^4(x_+)} 
    + \frac{1+ 2\cosh^2(x_-)}{\sinh^4(x_-)}\right) , \\
   \braket{ \varphi(x,\tau) \partial_x \theta(0,0)}
  =&  -\frac{T}{4 u} \left( \coth(x_+) - \coth(x_-) \right)  ,   \\
   \braket{ \varphi(x,\tau) \partial_x \varphi(0,0)}
  =&  -\frac{T K}{4 u} \left( \coth(x_+) + \coth(x_-) \right)    , \\  
   \braket{ \varphi(x,\tau) \partial^2_x \theta(0,0)}
  =&  -\frac{\pi T^2}{4 u^2} \left(\frac{1}{\sinh^2(x_+)} - \frac{1}{\sinh^2(x_-)} \right), \\
   \braket{ \left[ \varphi(x,\tau)-\varphi(0,0)\right]^2} 
  =& \frac{K}{2 \pi} \ln\left\lbrace \left( \frac{\beta u}{\pi a} \right)^2
    \sinh\left(\frac{\pi T}{u }( x- i u \tau  ) \right) \sinh\left(\frac{\pi T}{u }( x+ i u \tau )\right)                                            
    \right\rbrace \equiv \frac{K}{2 \pi} F(x,\tau)  .
\end{split}       
\end{align}
Here, we defined $x_{\pm} = \frac{\pi T}{u} \left\lbrace x \pm i [u \tau+ \text{sgn}(\tau) a]\right\rbrace $. The correlation functions for $\theta$ can be obtained from the ones above by the duality relation 
$ \sqrt{K} \varphi \, \to \,  \frac{1}{\sqrt{K}}\,  \theta$.  For later reference we also introduce the notation 
\begin{align}
   G_{\theta\varphi}^{(m)}(x-x',\tau-\tau') =& \braket{\partial^m_{x} \theta(x,\tau) \varphi(x',\tau') } \\
   G_{\theta\theta}^{(m,n)}(x-x',\tau-\tau') =&\braket{\partial^m_{x} \theta(x,\tau) \partial^n_{x'} \theta(x',\tau') }
\end{align}
These correlation functions will appear in the context of the \textit{ac} conductivity of a HLL in Appendix~\ref{App:Calculation of the conductivity of a disordered helical Luttinger liquid} and \ref{App:Calculation of the conductivity of a clean helical Luttinger liquid}. 

\subsection{Correlation functions containing exponentials of bosonic fields}

We often encounter correlation functions such as
\begin{equation}
   \braket{\theta'_{11} \theta'_{22} e^{2 i \sqrt{4 \pi} \left( \varphi_{33} -\varphi_{34} \right) }},
\end{equation}
where we denoted $\partial_x \theta(x,\tau) = \theta'(x,\tau)$ and $\theta(x_1,\tau_1) = \theta_{11}$. We can calculate them using the following trick:
\begin{align}
\begin{split} 
     \braket{\theta'_{11} \theta'_{22} e^{2 i \sqrt{4 \pi} \left( \varphi_{33} -\varphi_{34} \right) }} 
   =& \frac{1}{4(4 \pi)} \left. \partial_{I_1} \partial_{I_2} \right|_{I_1=I_2=0} \braket{e^{2 i \sqrt{4 \pi} 
     \left( \varphi_{33} -\varphi_{34} +I_1 \theta'_{11} -I_2 \theta'_{22}\right) }}\\
   =& \left\lbrace \braket{\theta'_{11} \theta'_{22}} - 16 \pi 
      \braket{\theta'_{11} \left(\varphi_{33} -\varphi_{34}   \right)}
      \braket{\theta'_{22} \left(\varphi_{33} -\varphi_{34}   \right)}   \right\rbrace  
      e^{-2 (4 \pi)  \braket{ \left( \varphi_{33} -\varphi_{34} \right)^2 }}.
\end{split}      
\end{align}
We employ this method of evaluating correlation functions containing exponentials of bosonic fields in the context of calculating the \textit{ac} conductivity in 
Appendix~\ref{App:Calculation of the conductivity of a disordered helical Luttinger liquid} and \ref{App:Calculation of the conductivity of a clean helical Luttinger liquid}.
\section{Calculation of the conductivity of a disordered helical Luttinger liquid}
\label{App:Calculation of the conductivity of a disordered helical Luttinger liquid}
In this Appendix we outline the calculation of \textit{ac} conductivity of a disordered HLL using full bosonization.
First, we expand the current-current correlation function to first order in impurity strength which yields
\begin{align}
   \braket{j^a(x,\tau) j^a(x',\tau')}  =& \braket{j^a_0(x,\tau) j^b_0(x',\tau')}_0 -  \braket{j^a_0(x,\tau) j^b_0(x',\tau') S_{\text{pert}}}_0 
                                           + 2 \braket{j^a_0(x,\tau) j^b_{\text{an,dis}}(x',\tau')}_0 + \mathcal{O}(D^2) .                                       
\end{align}
Here, we defined $ S_{\text{pert}} =  S_{\text{1P}}+ S_{\text{2P}}+ S_{\text{imp,b}}$. To first order the terms in $S_{\text{pert}}$ have to be diagonal in replica indices and therefore the replica limit is performed as
\begin{align}
      \begin{split} \frac{1}{R} \sum_{a,b,a'} \braket{j^a_0 j^b_0 S^{a'} } 
      =& \frac{1}{R} \sum_{a,b} \left( \sum_{a'}\Big|_{a'=b} \braket{j^a_0 j^b_0 S^{b} } 
      + \sum_{a'}\Big|_{a'\neq b}    \braket{j^a_0 j^b_0 S^{a'} }  \right)
      \stackrel{a \stackrel{!}{=} b}{=}  \frac{1}{R} \sum_{a=1}^R \left( \braket{j^a_0 j^a_0 S_{a} } + (R-1)
        \braket{j^a_0 j^a_0 } \braket{ S }  \right)\\
      \stackrel{R \to 0}{\to} & \braket{j_0 j_0 S} - \braket{j_0 j_0} \braket{S} 
      \equiv \braket{j_0 j_0 S}_c ,\end{split}
   \end{align} 
where we defined the connected average in the last equality.

We define the contributions linear in disorder strength as 
\begin{equation}
   \Sigma_1(x,x',\tau,\tau') = -  \braket{j^a_0(x,\tau) j^b_0(x',\tau') S_{\text{pert}}}_0 
+ 2 \braket{j^a_0(x,\tau) j^b_{\text{an,dis}}(x',\tau')}_0.
\end{equation}
Conductivity is then obtained by calculating the Fourier transform $\Sigma_1(k,k_n)$ and performing the limit 
\begin{equation}
   \sigma(\omega) = -\frac{i}{\omega} \left( \Sigma_1(k\to 0,i k_n\to \omega +i \delta) +\chi^{\text{dia}}(k,k_n) \right).
\end{equation}
We obtain
\begin{align}
    \Sigma^{\text{2P}}_1(k=0,k_n) =&   32  \frac{e^2 u^2 K^2}{k_n^2}  g_{\text{2P}} 
                                    \int_0^{\beta} \! \mathrm{d} \tau  \,  e^{- 4 K F(\tau)} \Bigl[ 1 - e^{i k_n \tau}\Bigr] , \\
    \Sigma_1^{\text{1P}}(k=0,k_n) =&   8  \frac{e^2 u^2 K^2}{k_n^2} g_{\text{1P}}  
                                    \int_0^{\beta} \! \mathrm{d} \tau \, G^{(2,2)}_{\theta \theta}(0,\tau) e^{- K F(\tau) }
                                    \Bigl[ 1 - e^{i k_n \tau}\Bigr], \\     
    \begin{split}                              
    \Sigma_1^{\text{imp,b}}(k=0,k_n) =&  8  \frac{e^2 u^2 K^2}{k_n^2}  g_{\text{imp,b}} 
                                    \int_0^{\beta} \! \mathrm{d} \tau \, \left\lbrace G^{(1,1)}_{\theta \theta}(0,\tau) - 4 \pi 
                                    \left[G^{(1)}_{\theta \varphi}(0,\tau) \right]^2 \right\rbrace e^{- K F(\tau) }
                                    \Bigl[ 1 - e^{i k_n \tau}\Bigr] \\      
                                  & +16 \frac{e^2 K u}{k_n} g_{\text{imp,b}} \int_0^{\beta} \! \mathrm{d} \tau \,
                                     G^{(1)}_{\theta\varphi}(0,\tau)e^{- K F(\tau) }  \Bigl[ 1 - e^{-i k_n \tau}\Bigr]
    \end{split}                                    
\end{align}
and
\begin{align}
   \chi^{\text{dia}}(k=0,k_n) =  -2 g_{\text{imp,b}} \frac{e^2}{\pi} \int \! \mathrm{d} \tau \, e^{-K F(\tau)} e^{i k_n \tau}   .
\end{align}
The conductivity due to 1P and 2P processes is then
\begin{align}
   \sigma^{\text{2P}}(\omega) =& 32 i \frac{e^2 u^2 K^2 }{\omega^3}    g_{\text{2P}} \left(\frac{\pi a T}{u} \right)^{8K} \mathcal{J}_{8K}(\omega,T), \\
   \sigma^{\text{1P}}(\omega) =& 8 i \frac{e^2 u^2 K}{\pi a^4 \omega^3} g_{\text{1P}}  \left(\frac{ \pi T}{u} \right)^{2K+4}
                                 \left(3 \mathcal{J}_{2K+4}(\omega,T)- 2  \mathcal{J}_{2K+2}(\omega,T) \right) ,
\end{align}
where we defined 
\begin{align}
\begin{split}
   \mathcal{J}_{2K}(\omega,T) =& \int_0^{\beta} \! \mathrm{d} \tau \, \frac{1-e^{i k_n \tau}}{\sin^{2K}(\pi \tau T)}
                          \Big|_{i k_n \to \omega+ i\delta}\\
                       =& \frac{2^{2K}}{T} \Gamma(1-2K) \left[\frac{1}{\Gamma^2(1-K)} - \frac{\sin(\pi K)}{\pi} 
                          \frac{\Gamma(K-i \frac{\omega}{2 \pi T})}{\Gamma(1- K-i \frac{\omega}{2 \pi T})} \right] . \label{J2K}
\end{split}                          
\end{align}
Here, $\Gamma(x)$ is the gamma function. These results appear in Eq.~(\ref{referAppE1}) and Eq.~(\ref{referAppE2}) of the main text.

In the case of backscattering off the impurity we obtain
\begin{align}
    \Sigma_1^{\text{imp,b}}(k=0,k_n) =& -4  e^2 K g_{\text{imp,b}} \left( \frac{\pi a T}{u} \right)^{2K} \left\lbrace
                                    \left(\frac{\pi T}{\omega} \right)^2 \left[ \left(2K +1\right) \mathcal{J}_{2K+2}(\omega,T) 
                                    -2 K \mathcal{J}_{2K}(\omega,T)   \right] 
                                    + 2 \frac{T}{\omega} \mathcal{L}_K(\omega,T) \right\rbrace ,   \\
     \chi^{\text{dia}}(k=0,k_n) =& \, 2 g_{\text{imp,b}} \frac{e^2}{\pi} \left( \frac{\pi a T}{u} \right)^{2K}  \frac{1}{\pi T} \sin(K \pi) 
                                   B(K-i\frac{\omega}{2 \pi T},1-2K)  .                          
\end{align}
Here, B(x,y) denotes the Euler beta function and we defined 
\begin{align}
\begin{split} 
   \mathcal{L}_K(\omega,T) =& \int \! \mathrm{d} \tau \, \frac{  1- e^{-i k_n \tau}}{\sin^{2K+1}(\pi T \tau)}\cos(\pi T \tau) \\
                       =& (-i) \sin(\pi K ) \frac{2^{2K}}{\pi T} \Bigl\lbrace 
                          \left[ B(K,-2K) -B(K-i\frac{\omega}{2 \pi T},-2K) \right]\\
                        & +\left[ B(K+1,-2K) -B(K+1-i\frac{\omega}{2 \pi T},-2K) \right] \Bigr\rbrace.
\end{split}                        
\end{align}
Adding the contributions yields $\Sigma_1^{\text{imp,b}}(k=0,k_n) + \chi^{\text{dia}}(k=0,k_n) =0$. Therefore, backscattering does not lead to a finite scattering time for any value of K to first order in $D_b$. This is discussed in Sec.~\ref{sec:bosonization:subsec:disordered case} of the main text.

\section{Calculation of the conductivity of a clean helical Luttinger liquid}
\label{App:Calculation of the conductivity of a clean helical Luttinger liquid}
The purpose of this Appendix is to outline the calculation of the \textit{ac} conductivity of a clean HLL using bosonization and the Kubo formula.

First, we expand the current-current correlation function to second order in interaction strength since the first order contribution vanishes due to the neutrality condition for vertex operators. This yields
\begin{align}
 \braket{j j} = \braket{j_0 j_0}_0 + \frac{1}{2}\braket{j_0 j_0 S_5^2}^c_0 -2 \braket{j_0 j_{\text{an,clean}} S_5}^c_0 
                +\braket{j_{\text{an,clean}} j_{\text{an,clean}}}_0 + \mathcal{O}(V^4).
\end{align}
Here, the connected averages appear due to the expansion of the denominator of the partition function.
As in the disordered case we define $\Sigma_2 \equiv \frac{1}{2}\braket{j_0 j_0 S_5^2}^c_0 -2 \braket{j_0 j_{\text{an,clean}} S_5}^c_0 
+\braket{j_{\text{an,clean}} j_{\text{an,clean}}}_0 $. Adding all the terms we are left with only one term contributing to the real part of the conductivity
\begin{align}
\begin{split} 
       \Sigma_2(x_3,x_4,\tau_3,\tau_4) 
   =&  \frac{1}{4} \frac{e^2 K^2 u^2}{\pi} \left(\frac{ 4 V k_F }{ \pi^{\frac{3}{2}} a k_0^2 } \right)^2
       \int \! \mathrm{d} x_1 \mathrm{d} \tau_1 \,  \int \! \mathrm{d} x_2 \mathrm{d} \tau_2 \, \\ 
    &  \times \braket{ \partial_{x_3} \theta(x_3,\tau_3)  \partial_{x_4} \theta(x_4,\tau_4) \partial_{x_1}^2 \theta(x_1,\tau_1)
       \partial_{x_2}^2 \theta(x_2,\tau_2) e^{i \sqrt{4 \pi} (\varphi(x_1,\tau_1) - \varphi(x_2,\tau_2))
       - 2 i k_F (x_1-x_2) }  }_0 
\end{split}       
\end{align} 
Using the methods outlined in Appendix~\ref{App:Formalism for conductivity calculation in the bosonized language} we obtain
\begin{align}
   \sigma(\omega) =&  \frac{i}{\omega^3} \frac{e^2 u^4 K^2}{h} \frac{2^6}{\pi^2}  
                      \left( \frac{V}{u}\right)^2 \left( \frac{k_F}{k_0}\right)^2 
                      \frac{1}{(a k_0)^2}  \mathcal{I}_K(\omega,T)  \label{AppF1}
\end{align}
Here, we defined
\begin{align}
\begin{split} 
   \mathcal{I}_K(\omega,T) 
   =& \left. \int \! \mathrm{d} x \int_0^{\beta} \mathrm{d} \tau \,  \left\lbrace G^{(2,2)}_{\theta \theta}(x,\tau)+ 4 \pi 
      \left[ G^{(2)}_{\theta \varphi}(x,\tau) \right]^2 \right\rbrace e^{- K F(x,\tau) } e^{2i k_F x} 
      \Bigl[ 1 - e^{i k_n \tau}\Bigr]\right|_{ik_n \to \omega +i \delta}  \\
   =& \, \frac{1}{(2 a)^{4} \pi u } \left(\frac{2 \pi T a}{u} \right)^{2K+4} \left( \frac{u}{\pi T}\right)^2 \sin(K \pi)
       \Bigl[ \frac{1}{K} \left\lbrace \mathcal{M}(\omega,-K,-K-2)+\mathcal{M}(\omega,-K-2,-K) \right\rbrace \\
    &     + \left(\frac{6}{K}+1 \right) \left\lbrace \mathcal{M}(\omega,-K,-K-4)+\mathcal{M}(\omega,-K-4,-K) \right\rbrace
          - 2 \mathcal{M}(\omega,-K-2,-K-2)  \Bigr] \label{IK} 
\end{split}\\
\begin{split}
   \mathcal{M}(\omega,\nu,\mu) =& B(-iS^0_- -\frac{\nu}{2},\nu+1)B(-iS_+^0-\frac{\mu}{2},\mu+1)
                                -B(-iS_- -\frac{\nu}{2},\nu+1)B(-iS_+ -\frac{\mu}{2},\mu+1) \label{AppF2}      
\end{split}                                           
\end{align}
and $S_{\pm} = \frac{\omega }{4 \pi T} \pm \frac{u k_F}{2 \pi T}$, $S_{\pm}^0 = S_{\pm}(\omega=0)$.
Eq.~(\ref{AppF1}) is Eq.~(\ref{refAppF}) of the main text.

\end{widetext}



\begin{thebibliography}{ee}
\bibitem{remark1} The $g_5$ term also appears in the model for a conventional LL and yields the corrections Eq.~(\ref{conductivityinteraction1}). However, since in a LL we always have $k_F \gg T$, the term can be safely neglected. This is not the case in the discussion of the HLL since we are naturally close to the Dirac point and thus have to incorporate the term in our treatment of transport properties.
\end{thebibliography}
\end{document}